\documentstyle[aps,multicol,epsf]{revtex}
\input epsf
\draft
\title{Atom focusing by far-detuned and resonant standing wave fields: Thin lens
regime.}
\author{J. L. Cohen$^{\dagger }$, B. Dubetsky$^{\ast }$, and P. R. Berman$^{\ast }$ 
\\ $^{\dagger }${\em Applied Physics Program, Department of Physics, University of\\
Michigan, Ann Arbor, MI 48109-1120}\\
$^{\ast }${\em Department of Physics, University of Michigan, Ann Arbor, MI\\
48109-1120}}
\date{\today }
\begin{document}
\maketitle
\begin{abstract}
The focusing of two-level atoms in a beam or trap after interacting with
both far-detuned and resonant standing wave fields in the thin lens and
paraxial approximations is considered theoretically. The thin lens
approximation is discussed quantitatively from a quantum perspective. Exact
quantum expressions for the Fourier components of the density (that include
all spherical aberration) are used to study the focusing numerically. The
following lens parameters and density profiles are calculated as functions
of the pulsed field area $\theta $: the position of the focal plane, peak
atomic density, atomic density pattern at the focus, focal spot size, depth
of focus, and background density. The lens parameters are compared to
asymptotic, analytical results derived from a scalar diffraction theory for
which spherical aberration is small but non-negligible ($\theta \gg 1$).
Within the diffraction theory analytical expressions show that the focused
atoms in the far detuned case have an approximately constant background
density $0.5(1-0.635\theta ^{-\,1/2})$ while the peak density behaves as $%
3.83\theta ^{1/2}$, the focal distance or time as $\theta ^{-1}(1+1.27\theta
^{-\,1/2})$, the focal spot size as $0.744\theta ^{-3/4}$, and the depth of
focus as $1.91\theta ^{-\,3/2}$. Focusing by the resonant standing wave
field leads to similar results. However, resonant focusing is also
accompanied by a new effect, a Rabi-like oscillation of the atom density.
For the far-detuned lens, chromatic aberration caused by the longitudinal
velocity distribution in an atom beam is studied quantitatively with the
exact Fourier results. Similarly, the degradation of the focus that results
from angular divergence in beams or thermal velocity distributions in traps
is studied quantitatively with the exact Fourier method and understood
analytically using the asymptotic results. Overall, we show that strong thin
lens focusing is possible with modest laser powers and with atomic beam
characteristics that are currently achievable in the laboratory.
\end{abstract}


\begin{multicols}{2}
\narrowtext

\section{Introduction}

Evolving from the initial experiments\cite{3,2}, the study of atom focusing
using standing wave (SW) light fields has developed into a broad area of
research in atom optics. A standing wave field acts as a lens as a result of
a spatially-dependent light shift, focusing atoms into a periodic set of
lines or dots having widths of the order of tens of nanometers and distanced
from one another by $\lambda /2$, where $\lambda $ is the wave length of the
light. Focusing by thick lenses and atomic deposition on a substrate have
been observed for beams of Na\cite{3,5,4,6,6p,61}, metastable He \cite{2},
and Cr\cite{7,4p,8}. The latest achievements in this area are summarized in
a recent review article\cite{81}. Perhaps even more significantly,
lithographic techniques have been developed to create nanosurfaces in
semiconductors and metals using metastable atoms as a pattern template for
selective etching\cite{8a}. Similarly, cold, trapped atoms can be subjected
to SW light pulses\cite{12,13}, creating a periodic wave packet which will
focus along the field propagation direction at specific times following the
atom-field interaction.

Previous experiments with atomic beams have been carried out in a thick lens
regime for which the atoms focus within the laser beam. Since the substrate
surface is typically placed near the plane of peak intensity of the thick SW
lens and mechanically fixed to the retroreflecting mirror, in some sense
these experiments are easier to set up when compared to thin lens
experiments. The classical and quantum motion through a thick lens have been
numerically simulated and compared to experimental data in Refs.\cite
{61,82,83,83b,83bb}. The classical motion and lens parameters after thick
and thin lenses have also been studied in detail within a ray optics
formalism\cite{84}. For that work the lens characteristics and aberrations
beyond the parabolic lens approximation were simulated numerically as were
the effects of different atomic velocity distributions. However, a
comprehensive theoretical study of the thin SW lens from a quantum
perspective that details corrections to the lens parameters as a result of
spherical aberration, chromatic aberration, and angular beam divergence has
not appeared previously.

In this article we stress the flexibility and validity of a thin lens
approach, where the atoms are detected or used for lithography after the
interaction with the field. We derive exact and computationally expedient
results for atomic matter waves by quantizing the center-of-mass motion from
the outset, a process that naturally accounts for spherical aberration. Our
hope is to stimulate interest in thin lens experiments by introducing a
straightforward, yet rigorous theoretical framework for the problem and by
deriving analytical results for lens parameters to facilitate lithographic
and atom optical configurations. The thin SW lens creates high contrast
periodic structures in free flight and opens up numerous possibilities for
Fourier atom optics to image and manipulate modulated matter waves. This
approach is highly desirable, especially when coupled with advances in
Bose-Einstein condensate (BEC) atom laser sources\cite{13,BEC} with large
fluxes and narrow velocity distributions. To our knowledge similar studies
have been carried out only numerically using a Fresnel diffraction theory in
the coordinate representation for a ''doughnut''-mode optical field\cite{8pp}%
, standing wave field\cite{85}, and conical lens\cite{86}.

Rather than using the diffraction theory simply for numerical propagation of
the matter waves, we extend its usefulness by deriving approximate
analytical expressions for the SW field lens parameters. The
Fresnel-Kirchhoff integral, incorporating the lowest-order spherical
aberration, gives a relatively simple form for the atomic wave function near
the focal plane and SW intensity extrema. Asymptotically for large
atom-field pulse areas, this integral leads to a universal atomic density
profile, revealing the behavior of the focal distance, peak atomic density,
focal spot size [half-width at half-maximum (HWHM) of the atom density at
the focus], depth of focus, and background density as a function of the
pulse area. The analytical asymptotic results can then be compared to either
numerical asymptotic approximations of the Fresnel integral or to the exact
atomic distributions and lens parameters, most easily evaluated using a
Fourier approach.

The time evolution of a free particle's wave function, when subject to a
spatially periodic initial condition such as that imposed on an atom beam
after passing through a SW field or microfabricated grating\cite{12,87,rev},
can be expressed in terms of a Fourier expansion in the (plane wave) spatial
harmonics of the modulation period. As such, the results we derive are
closely related to the canon of research on periodic light optics\cite{86a}.
In the thin lens approximation the Fourier components can be calculated
exactly for the wave function and density and then superposed to evaluate
the atomic spatial distribution: for this approach the calculations
effectively take place in a momentum representation, whereas a numerical
integration over the free particle propagator gives results in a coordinate
representation. The expressions for the density's Fourier components that we
use below were obtained recently for far-detuned \cite{411} and resonant\cite
{412} SW fields. To our knowledge this exact solution has not been used
previously to examine periodic focusing parameters, nor to compare with
approximate solutions of focused waves. However, the Fourier components and
the degradation of this type of focusing have been proposed recently as a
way to study the onset of many-body effects in condensates\cite{Jayson1}.

The Fourier approach has several advantages. First, the time dependence of
the Fourier coefficients directly reveals the quantum nature of the periodic
atomic density as it varies explicitly with the atomic recoil energy of
integer pairs of photons. Second, summation of the Fourier series leads to
fast convergence of the Fresnel solutions to arbitrary accuracy for all
times when compared to the numerical integration of the full
Fresnel-Kirchhoff integral. With this Fourier method for example, we can
reproduce the main Fresnel diffraction results of Ref.\cite{85} without
having to integrate the partial differential equation or equivalent integral
equation on a lattice. Furthermore, the Fourier method can be employed for
the general case of typical atomic beam experiments, where the incident beam
or trapped atoms are most accurately specified by a density matrix in
momentum space, such as a thermal velocity distribution. The implications
are that the spatial extent of the atomic distribution is much larger than
the wavelength of the standing wave and the wave function is not transform
limited (except perhaps for the case of a condensate), so that the focal
densities are conveniently found as a function of this initial density
matrix. As a direct result, while the failure to account for finite beam
size effects with the Fourier series limits our calculations to the
near-field and Fresnel diffraction regimes, it is relatively unimportant for
atom focusing immediately following the SW pulse. We can then calculate the
effects of chromatic aberration and angular beam divergence by averaging the
exact Fourier solution over any longitudinal and transverse atomic velocity
distributions.

In addition to lithographic detection schemes, one can detect atomic
distributions modulated by SW fields by backscattering a pulsed traveling
wave probe off of the density either (a) at a certain distance from the
grating for the beam case or (b) after a certain time for the atoms released
from a trap. The scattered signal is sensitive to the lowest-order Fourier
component\cite{41} of the modulated density. Previously, the backscattered
signal from a cold vapor subjected to strong SW pulses was observed in Rb in
a ground state echo experiment\cite{12}. The excellent agreement between the
Fourier theory and experiment that was achieved in that case insures that
our theoretical approach can be applied to the atom focusing problem as well.

A new regime of atom focusing arises if the SW field is resonant with the
transition between a ground and excited state and the time of interaction is
less than the excited state lifetime. After propagating from the SW field
for a time just long enough for the excited state to decay, the atom spatial
distribution consists of two parts\cite{412}: the stimulated density, caused
by direct amplitude modulation of the ground state wave function by the
field, and the spontaneous density, produced by excitation to and decay from
the excited state. As a result of the periodic Talbot revival, or
self-imaging \cite{86a,22,23,24}, of the atomic spatial distribution, free
evolution of the stimulated part leads to atom focusing not only at the
focal distance $L_{f},$ but also at the distances $jL_{T}/2\pm L_{f}$, where 
$j$ is a positive integer, $L_{T}=\lambda ^{2}/2\lambda _{dB}$ is a Talbot
length, and $\lambda _{dB}$ is the atom de Broglie wavelength. In contrast
to the stimulated part,\ the spontaneous modulation decays at a distance on
the order of $L_{T}$ as a result of the Doppler shift associated with
spontaneous emission. At a sufficiently large distance, the modulated
spontaneous part disappears, and one can observe and analyze atom focusing
by a resonant SW field. This analysis then reveals a new effect, a Rabi-like
oscillation of the focal parameters with the atom-field interaction strength
that arises from an interference between different components of the matter
wave.

This article is arranged as follows. In Section II we consider the focusing
by a far-detuned field acting as a phase grating. Focusing by a resonant
field acting as an amplitude grating is studied in Section III. Section IV
is devoted to the influence of chromatic aberration (longitudinal velocity
distributions) and angular beam divergence (transverse velocity
distributions). The quantitative results and illustrative examples are
discussed in Section V.

\section{Far-detuned standing wave lens}

Atom optical experiments can operate in the spatial or time domain. For
example, in the time domain a vapor of cold atoms interacts with one or more
radiation pulses forming spatial gratings in the $x$ direction.
Correspondingly, in the spatial domain an atom beam traverses one or more
optical elements or interaction regions. For a monovelocity beam propagating
along the $z$ axis with longitudinal velocity $U$ and with the optical
elements aligned in the $x$ direction, the spatial domain configurations can
be analyzed in the time domain if calculations are performed in the atomic
rest frame moving with velocity $U$. In this frame the optical elements
appear as interaction pulses. As a result, our calculations are restricted
to the time domain without loss of generality (and are adapted to account
for a longitudinal velocity distribution in Section IV). We consider the
focusing effects of a single pulse.

The atom optical elements couple to the center-of-mass degrees of freedom of
an atom with mass $M$ and de Broglie wavelength $\lambda _{dB}=2\pi \hbar
/(MU)$. When the SW laser field, 
\begin{equation}
E\left( x,t\right) =Ee^{-i\Omega t}g(t)\cos \left( kx\right) +c.c.,
\label{1}
\end{equation}
drives the atomic transition between the ground state $\left| g\right\rangle 
$ and the excited state $\left| e\right\rangle $, the Schr\"{o}dinger
equation, $i\hbar \partial {\bf \Psi }/\partial t=H{\bf \Psi }$, governs the
motion. The Hamiltonian in the rotating-wave approximation is 
\begin{eqnarray}
H &=&\frac{p_{x}^{2}}{2M}+\hbar \omega \left| e\right\rangle \langle e| 
\nonumber \\
&&+2\hbar \chi g(t)\cos \left( kx\right) \left( e^{-i\Omega t}\sigma
_{+}+e^{i\Omega t}\sigma _{-}\right) ,  \label{1a}
\end{eqnarray}
where 
\begin{equation}
{\bf \Psi }=\left( 
\begin{tabular}{l}
$\psi _{e}(x,t)$ \\ 
$\psi (x,t)$%
\end{tabular}
\right)   \label{1b}
\end{equation}
is the two-state wave function for the ground ($\psi (x)$) and excited ($%
\psi _{e}(x)$) states, $p_{x}$ is the center-of-mass momentum operator, $%
\omega $ is the atomic transition frequency, $\Omega ,k=q/2=2\pi /\lambda ,$
and $g\left( t\right) $ are the frequency, propagation wave vector, and
pulse envelope function (centered at $t=0$, having peak value of unity and
duration on the order of $\tau ),$ respectively, of the field, $\sigma
_{+}(\sigma _{-})$ is the atomic raising (lowering) operator associated with
the transition $\left| g\right\rangle \rightarrow \left| e\right\rangle $ $%
(\left| e\right\rangle \rightarrow \left| g\right\rangle ),$ and $\chi =-\mu
E/2\hbar $ is a Rabi frequency for the $\left| g\right\rangle \rightarrow
\left| e\right\rangle $ transition with dipole matrix element $\mu $. We
have assumed for simplicity that the Rabi frequency $\chi $ is real.

In the spatial domain this Hamiltonian is written in the atomic rest frame
for the slowly-varying wave function ${\bf \Psi }$ in the paraxial optics
limit, 
\begin{equation}
MU\gg \left\langle p_{x}\right\rangle ,\sqrt{\left\langle
p_{x}^{2}\right\rangle -\left\langle p_{x}\right\rangle ^{2}};
\end{equation}
for $t>0$ the distance from the interaction region in the lab frame is $L=Ut$%
. In the time domain a paraxial approximation is unnecessary, and ${\bf \Psi 
}$ is simply the wave function. Experimentally\cite{6p}, for focusing it has
proven advantageous to optically pump the atoms into an initial ground state
with magnetic quantum number $m_{j}=\pm j$ and use circularly polarized
fields to avoid multiple Rabi frequencies in the state dynamics, justifying
our two-level approximation.

Pure phase modulation\ of the atomic ground state during the interaction
occurs in the far detuned case 
\begin{equation}
\Delta \gg \max \left( \Gamma ,\tau ^{-1}\right) ,  \label{a4}
\end{equation}
where $\Delta =\Omega -\omega $ is the atom-field detuning, and $\Gamma $ is
the excited state decay rate. Often, a steady-state light shift potential,
derived from the Hamiltonian (\ref{1a}), has been used as an effective
atom-field interaction\cite{25}, 
\begin{equation}
\frac{\hbar \Delta }{2}\ln \left[ 1+\frac{8\left| \chi \right|
^{2}g^{2}(t)\cos ^{2}\left( kx\right) }{(\Gamma /2)^{2}+\Delta ^{2}}\right] .
\label{a4'}
\end{equation}
While this potential is important for smaller detunings and/or larger
intensities, it is strictly valid only for $\Gamma \tau \gg 1$. On the other
hand, the dressed state potential, 
\begin{equation}
\frac{\hbar \Delta }{2}\sqrt{1+16\left| \chi \right| ^{2}g^{2}(t)\cos
^{2}\left( kx\right) /\Delta ^{2}},  \label{a4''}
\end{equation}
neglects spontaneous emission and assumes that the atom in its ground state
adiabatically evolves into one of the two dressed states. For our work we
follow the experimental findings of Natarajan and coworkers\cite{6p}. They
show that focusing is improved by taking a short interaction time and a
large detuning relative to the Rabi frequency and decay rate. In our case
these conditions avoid spontaneous emission (diffusive aberration\cite{83b})
both during and after the atom-field interaction. In this limit the two
potentials, Eqs. (\ref{a4'}) and (\ref{a4''}), reduce to the same effective
potential, ignoring spatially independent energy terms. The Hamiltonian in a
field interaction representation for the ground state wave function $\psi
\left( x,t\right) $ after adiabatically eliminating the excited state for $%
\Delta \gg 2\left| \chi \right| $ and $4\Gamma \tau |\chi |^{2}/\Delta
^{2}\ll 1$ ($\psi _{e}(x,t)\sim 2\chi /\Delta \simeq 0$) is 
\begin{equation}
H=\frac{p_{x}^{2}}{2M}+\frac{2\hbar \left| \chi \right| ^{2}}{\Delta }%
g^{2}(t)\cos (qx)\text{.}  \label{a5b}
\end{equation}

If the incident atom wave function is uniform $\left[ \psi \left(
x,t=0^{-}\right) =1\right] ,$ then just after a single interaction the wave
function is given by 
\begin{mathletters}
\label{a6''}
\begin{eqnarray}
\psi \left( x,t=0^{+}\right) &=&\exp \left[ i\left( \theta /2\right) \cos
\left( qx\right) \right]  \label{a6} \\
&=&\sum_{n=-\infty }^{\infty }i^{n}J_{n}(\theta /2)e^{inqx},  \label{a6'}
\end{eqnarray}
where 
\end{mathletters}
\begin{equation}
\theta =-\left( \frac{4\left| \chi \right| ^{2}}{\Delta }\right)
\int_{-\infty }^{\infty }dtg^{2}\left( t\right)  \label{2}
\end{equation}
is an effective pulse area for the far-detuned atom-field interaction, and $%
J_{n}$ is a Bessel function of order $n$. Equation (\ref{a6}) is valid in
the Raman-Nath\cite{RN} or thin lens approximation, for which the $%
p_{x}^{2}/(2M)$ term is ignored during the interaction so that the field
acts as a standing wave phase grating for the atoms. A standard condition
given for the Raman-Nath approximation for thin lens focusing is\cite{83b,85}
\begin{equation}
\left| \theta \right| \omega _{q}\tau /2\ll 1\text{.}  \label{3}
\end{equation}
The two-photon recoil frequency is 
\begin{equation}
\omega _{q}=\frac{\hbar q^{2}}{2M}=2\pi \frac{L_{T}}{U}\text{,}  \label{4}
\end{equation}
and $L_{T}=\lambda ^{2}/(2\lambda _{dB})$ is the Talbot distance for the
atom beam. The momentum eigenstates, $\exp [inqx],$ which are coherently
superposed to form this wave function, each have the free particle energy $%
n^{2}\hbar \omega _{q}$.

\subsection{Thin versus thick lens regimes}

Before we proceed, a clarification is needed to emphasize the differences
between the thin lens (Raman-Nath) and thick lens regimes of the SW atom
lens. Some of the quantitative differences have been examined by Henkel and
coworkers\cite{101} in relation to Fraunhofer diffraction of atoms and from
a ray optics (classical) perspective by McClelland for focusing\cite{84}.
For focusing considerations we assume square pulses, $g(t)=1$ for $-\tau
/2\leq t\leq \tau /2$ and zero otherwise, in order to obtain quantitative
results. \ We also require 
\begin{equation}
\omega _{q}\tau \ll 1,\left| \theta \right| \gg 1\text{.}  \label{6}
\end{equation}
Conditions (\ref{6}) are necessary for high contrast, thin lens atom
focusing, where we are interested in the atoms after propagating through the
lens. (The pulse shape will affect the coefficients of our results, not the
scaling with $\theta $ and $\omega _{q}\tau $.)

The Raman-Nath regime leading to condition (\ref{3}) is normally defined as
an interaction for which the average kinetic energy gained by the atoms
remains much smaller than the interaction strength coupling the momentum
components, $\left\langle p^{2}(t)/(2M)\right\rangle \ll \hbar \left| \chi
\right| ^{2}/|\Delta |$ in the square pulse case. If this condition is
violated while (\ref{6}) holds, the lens is thick, and the atoms can focus
within the interaction region near the time $t\simeq -\tau /2+\pi (\left|
\chi \right| ^{2}\omega _{q}/|\Delta |)^{-1/2}/4<\tau /2$. If the atoms have
not focused completely by the end of the pulse, they will exit the
interaction region amplitude modulated. To lowest order in $\theta \omega
_{q}\tau $, the amplitude correction which multiplies the wave function (\ref
{a6}) immediately after the square pulse can be calculated to be $\exp
[\theta \omega _{q}\tau \cos \left( qx\right) /4]$.

However, the atoms also acquire an additional, spatially-modulated phase
shift, $\theta ^{2}\omega _{q}\tau \cos \left( 2qx\right) /24$ to lowest
order in $\theta ^{2}\omega _{q}\tau $. This is called the WKB\ correction
by Henkel and coworkers\cite{101} and has its semiclassical origin in the
harmonic motion of the atoms during the interaction. From a quantum
perspective these amplitude and phase changes are caused simply by the
kinetic energy acquired during the interaction. Heuristically, using the
wave function which is evolving in the SW field, $\psi \left( x,t\right)
=\exp [-i2\left| \chi \right| ^{2}t\cos \left( qx\right) /\Delta ],$ the
correction which multiplies $\psi \left( x,\tau /2\right) $ can be written
as 
\begin{eqnarray}
&&\left. \exp [\frac{-i}{\hbar }\int_{-\tau /2}^{\tau /2}dt\psi ^{\ast
}\left( x,t\right) \frac{p^{2}}{2M}\psi \left( x,t\right) ]=\right.  \\
&&\exp [\theta \omega _{q}\tau \cos \left( qx\right) /4-i\omega _{q}\tau
\theta ^{2}(1-\cos (2qx))/24],
\end{eqnarray}
giving the rigorously correct result. If we are interested in far-field
diffraction, the momentum state wave function can change significantly from
the spectrum of Eq. (\ref{a6'}) if $\theta ^{2}\omega _{q}\tau /24\gtrsim 1$
even if condition (\ref{3}) is satisfied.

In the Fresnel focal region of interest here, while the spatially-dependent
phase shift is crucial in determining the thin lens properties of the SW
field, we require $\left| \theta \right| /2\gg \theta ^{2}\omega _{q}\tau /6$
to assure the dominance of the Raman-Nath wave function (\ref{a6}) near the
focus. In other words a corrected Raman-Nath condition, 
\begin{equation}
\left| \theta \right| \omega _{q}\tau /3\ll 1\text{,}  \label{RN}
\end{equation}
is sufficient to observe thin lens focusing, but the additional thin lens
condition, $\theta ^{2}\omega _{q}\tau \cos \left( 2qx\right) /24\ll 1$, may
be necessary to ignore corrections to Eq. (\ref{a6}) for other observables,
like the far-field diffraction pattern or the time-dependent behavior of
Fourier components of the density in echo configurations. Putting the
results together, the corrected wave function to lowest order in $\theta
^{2}\omega _{q}\tau ,$ $\theta \omega _{q}\tau ,$ 
\begin{eqnarray}
&&\left. \psi \left( x,\tau /2\right) =\exp \left[ i\left( \theta /2\right)
\cos \left( qx\right) \right] \right.  \\
&&\times \exp \left[ i\theta ^{2}\omega _{q}\tau \cos \left( 2qx\right) /24%
\right] \exp \left[ \theta \omega _{q}\tau \cos \left( qx\right) /4\right] ,
\end{eqnarray}
could be incorporated into the theoretical work below if necessary. We have
verified these quantitative results using a Crank-Nicholson technique to
integrate the Schr\"{o}dinger equation for the far-detuned SW Hamiltonian (%
\ref{a5b}) numerically on a lattice\cite{koonin}.

\subsection{Results for far-detuned, thin lens focusing}

Returning to Eqs. (\ref{a6''}) as the initial condition for the free motion,
since only the wave function phase has been changed during the interaction,
the total atom density, 
\begin{equation}
\rho \left( x,t\right) =\left| \psi \left( x,t\right) \right| ^{2}+\left|
\psi _{e}\left( x,t\right) \right| ^{2}\simeq \left| \psi \left( x,t\right)
\right| ^{2},
\end{equation}
is initially uniform. As different Fourier components of the wave function
acquire different phase shifts, $\varphi _{n}(t)=n^{2}\omega _{q}t$, during
the free evolution, the atom density for $t>0$ becomes spatially modulated.
The period of the spatial modulation is equal to $\lambda /2=2\pi /q$.
Transferring to the dimensionless variables, 
\begin{equation}
x\rightarrow qx,\text{ }t\rightarrow \omega _{q}t,  \label{a8}
\end{equation}
for $t>0$ one finds\cite{86a,411} 
\begin{mathletters}
\label{5'}
\begin{eqnarray}
\psi \left( x,t\right) &=&\sum_{n=-\infty }^{\infty }i^{n}J_{n}\left[ \theta
/2\right] e^{i(nx-n^{2}t)},  \label{5a} \\
\rho \left( x,t\right) &=&\sum_{n=-\infty }^{\infty }J_{n}\left[ \theta \sin
\left( nt\right) \right] e^{inx}.  \label{5b}
\end{eqnarray}
This result, exact in the thin lens approximation, is used below for the
numerical study of atom focusing. We refer to the exact calculations by this
Fourier technique as Method 1 in the text and figures that follow.

Arbitrarily precise values of the focal time $t_{f}$ (giving the focal plane
position for the beam, $L_{f}=Ut_{f}$) for a given $\theta $ are defined by
the first maximum of the density along $x=0$, $\rho \left( 0,t\right) $, as
a function of $\theta $ using Eq. (\ref{5b}). Given the focal time, we can
further characterize the lens by the density profile at the focus $\rho
\left( x,t_{f}\right) $ and its peak $\rho \left( 0,t_{f}\right) $, the spot
size of the focus $w$, the depth of focus $\Delta t$ , and the background
density $\rho \left( \pi ,t_{f}\right) $\cite{focal}. The spot size $w$
(HWHM of the density profile $\rho (x,t_{f}),$ where $\rho \left(
0,t_{f}\right) $ is the maximum) can be defined implicitly as the smallest
positive root of the equation 
\end{mathletters}
\begin{equation}
\rho (w,t_{f})=\,^{1}/_{2}\rho (0,t_{f})  \label{8}
\end{equation}
Similarly, the depth of focus (confocal parameter) can be defined using Eq. (%
\ref{5b}) as the time window, 
\begin{equation}
\Delta t=t_{+}-t_{-}\text{ for }t_{-}<t_{f}<t_{+}\text{,}  \label{depth1}
\end{equation}
within which the density along $x=0$ rises from half its peak value at $%
t_{-} $ to its peak value at $t_{f}$ and back again at $t_{+}$, given
implicitly by 
\begin{equation}
\rho (0,t_{\pm })=\,^{1}/_{2}\rho (0,t_{f})\text{.}  \label{depth2}
\end{equation}
This region is not symmetric with respect to the focal time $t_{f}$ as a
result of the spherical aberration of the lens. (Note that the choice of $%
\theta $ as positive is unimportant even though it implies a red detuning of
the field. The $\theta <0$ case is identical but shifted in $x$ by $\pi $.)

Several of these exact lens parameters, as calculated by Method 1, are
compared to the approximate diffraction theory below. Equations (\ref{5b})
will be modified to account for finite beam divergence and chromatic
aberration in Sec. IV; by its nature the Fourier method includes spherical
aberration (anharmonicity in the SW potential) to all orders in the lens
curvature. Although Eqs. (\ref{5'}) are exact, they offer no transparent
possibilities for obtaining analytical forms for the lens parameters as
functions of $\theta .$ In contrast, a diffraction theory can be used to
find approximate, asymptotic expressions ($\theta \gg 1)$ for the $\theta $%
-dependences of the lens parameters.

Rewriting Eq. (\ref{5a}) as an integral in the coordinate representation, we
can express the wave function as 
\begin{equation}
\psi \left( x,t\right) =\int_{-\infty }^{\infty }dx^{\prime }G\left(
x-x^{\prime },t-t^{\prime }\right) \psi \left( x^{\prime },t^{\prime
}\right) ,  \label{a7}
\end{equation}
where the Fresnel-Kirchhoff propagator of the free atom motion is given in
the dimensionless variables (\ref{a8}) by 
\begin{equation}
G\left( x,t\right) =\left( 4\pi it\right) ^{-\,^{1}/_{2}}\exp \left(
ix^{2}/4t\right) \text{.}  \label{a9}
\end{equation}
The integrand wave function is taken as $\psi \left( x^{\prime },t^{\prime
}=0\right) =\exp [i(\theta /2)\cos (x^{\prime })]$. In this form we are
first interested in the wave function behavior near the focal points, $%
x_{m}=2\pi m,$ for integer $m$ and times $t$ immediately following the
interaction. The relevant time scale for focusing will become apparent
shortly. When $\theta \gg 1$, the main contribution to the integral (\ref{a7}%
) near $x_{m}$ at these times arises from small values of $|x^{\prime
}-x_{m}|$, where the potential in Eq. (\ref{a5b}) is nearly harmonic. We
choose to expand around the point $x^{\prime }=0$ ($m=0$) and downplay the
periodicity of the wave function for the diffraction theory. Replacing $\cos
\left( x^{\prime }\right) $ by $1-x^{\prime 2}{}/2,$ the atom density at the
center, 
\begin{equation}
\rho (0,t)\simeq \left( 1-t\theta \right) ^{-1},  \label{a101}
\end{equation}
contains a singularity at $t=\theta ^{-1}$ which determines the approximate
position of the focal plane.

To obtain a finite value for the atomic density, spherical aberration ({\it %
i.e.,} anharmonic terms in the potential) must be considered. Expanding $%
\cos \left( x^{\prime }\right) $ to the $x^{\prime 4}$ term, omitting the
phase factor $\exp \left( ix^{2}/4t\right) ,$ and choosing the scaled
position $\xi =\,^{1}/_{2}\left( \theta /3\right) ^{^{1}/_{4}}x^{\prime }$
as an integration variable, we find that the asymptotic wave function in the
vicinity of $x=0$ is 
\begin{mathletters}
\label{a11}
\begin{eqnarray}
\psi \left( x,t\right) &\sim &\left( 3/\theta \right) ^{^{1}/_{4}}\left( \pi
it\right) ^{-\,^{1}/_{2}}\exp \left( i\theta /2\right) f\left( \tilde{x},%
\tilde{\omega}\right) ,  \label{a11a} \\
f\left( \tilde{x},\tilde{\omega}\right) &=&\int_{-\infty }^{\infty }d\xi
\exp \left( -i\tilde{x}\xi +i\tilde{\omega}\xi ^{2}+i\xi ^{4}\right) ,
\label{a11b} \\
\tilde{x} &=&\left( 3/\theta \right) ^{^{1}/_{4}}x/t,  \label{a11c} \\
\tilde{\omega} &=&\left( 3/\theta \right) ^{\,^{1}/_{2}}\left( t^{-1}-\theta
\right) .  \label{a11d}
\end{eqnarray}
The main contribution to the wave function comes from the region\ where the
integrand phase originating from the anharmonic term is of the order of
unity $\left( \left| \xi _{\max }\right| \sim 1\text{ or }\left|
x_{\max }^{\prime }\right| \sim \theta ^{-\,^{1}/_{4}}\right) .$ To
determine the validity of Eqs. (\ref{a11}), the next anharmonic term ($\sim
x^{\prime 6}$) must produce a small addition to the integrand phase in this
region, implying $\theta x_{\max }^{\prime 6}\ll 1,$ or 
\end{mathletters}
\begin{equation}
\theta ^{-\,^{1}/_{2}}\ll 1.  \label{a111}
\end{equation}
This condition defines the proper asymptotic limit - a wave function
correction of relative weight $\theta ^{-\,^{1}/_{2}}$ in Eq. (\ref{a11a})
could be included to increase the calculation's accuracy. We assume
condition (\ref{a111}) holds.

\mbox{$>$}%
From Eqs. (\ref{a11}) the time-dependent density, 
\begin{mathletters}
\label{a12}
\begin{eqnarray}
\rho \left( x,t\right)  &=&\left| \psi \left( x,t\right) \right| ^{2}\sim
\left( 3/\theta \right) ^{^{1}/_{2}}\left( \pi t\right) ^{-\,1}F\left( 
\tilde{x},\tilde{\omega}\right)   \label{a12a} \\
F\left( \tilde{x},\tilde{\omega}\right)  &=&\left| f\left( \tilde{x},\tilde{%
\omega}\right) \right| ^{2},  \label{a12b}
\end{eqnarray}
contains both the slowly-varying $t^{-1}$ dependence and the sharp
dependence of $F\left( \tilde{x},\tilde{\omega}\right) $ in the vicinity of $%
t=\theta ^{-1}$. At the focal center ($\tilde{x}=0$), the integral (\ref
{a11b}) can be expressed analytically (see Ref. \cite{Ryzhik}, No. 3.696)
through fractional Bessel functions as 
\end{mathletters}
\begin{eqnarray}
&&\left. f\left( 0,\tilde{\omega}\right) =2^{-\,^{3}/_{2}}\pi \left| \tilde{%
\omega}\right| ^{^{1}/_{2}}\exp \left[ -i\left( \tilde{\omega}^{2}-\pi
\right) /8\right] \right.   \nonumber \\
&&\times \left[ J_{-\,^{1}/_{4}}\left( \tilde{\omega}^{2}/8\right)
+i^{^{3}/_{2}}sign\left( \tilde{\omega}\right) J_{^{1}/_{4}}\left( \tilde{%
\omega}^{2}/8\right) \right] ,  \label{a13}
\end{eqnarray}
giving an analytical asymptotic expression for the density as a function of
time along $x=0,$%
\begin{equation}
\rho \left( 0,t\right) \sim \left( 3/\theta \right) ^{^{1}/_{2}}\left( \pi
t\right) ^{-\,1}\left| f\left( 0,\tilde{\omega}\right) \right| ^{2}\text{.}
\label{a13a}
\end{equation}

The lens parameters in this asymptotic diffraction theory are defined in the
same way as they were for the exact Fourier theory above. They are found
graphically or numerically by evaluating Eqs. (\ref{a11}), (\ref{a12}), and (%
\ref{a13a}). The focal time $t_{f}$ for a given $\theta $ (which is not
given by $t=\theta ^{-1})$ can be found as the first maximum of $\rho \left(
0,t\right) $ from Eq. (\ref{a13a}). Using this focal time $t_{f}$, we
further characterize the SW lens by the peak density $\rho (0,t_{f})$
evaluating Eq. (\ref{a13a}), by the density profile at the focus $\rho
(x,t_{f})$ found from Eqs. (\ref{a11}) and (\ref{a12}), by the spot size $w$
(HWHM of the density profile $\rho (x,t_{f})$), and by the depth of focus $%
\Delta t$ found from Eqs. (\ref{a12a}) and (\ref{a13a}).

Calculations performed with these asymptotic diffraction results, only
restricting $x$ to be near $x_{m}=2\pi m$, are referred to as Method 2. In
addition to the Method 2 asymptotic results, we can derive expressions which
are valid both for $x\simeq x_{m}$ and for the time restricted to be near
the focus, $t\simeq \theta ^{-1},$ namely within the depth of focus. This
further approximation, which is referred to as Method 3, leads to analytical
expressions for the lens parameters in the asymptotic limit.

\begin{figure}[tb!]
\centering
\begin{minipage}{8.0cm}
\epsfxsize= 8 cm \epsfysize= 11.95 cm \epsfbox{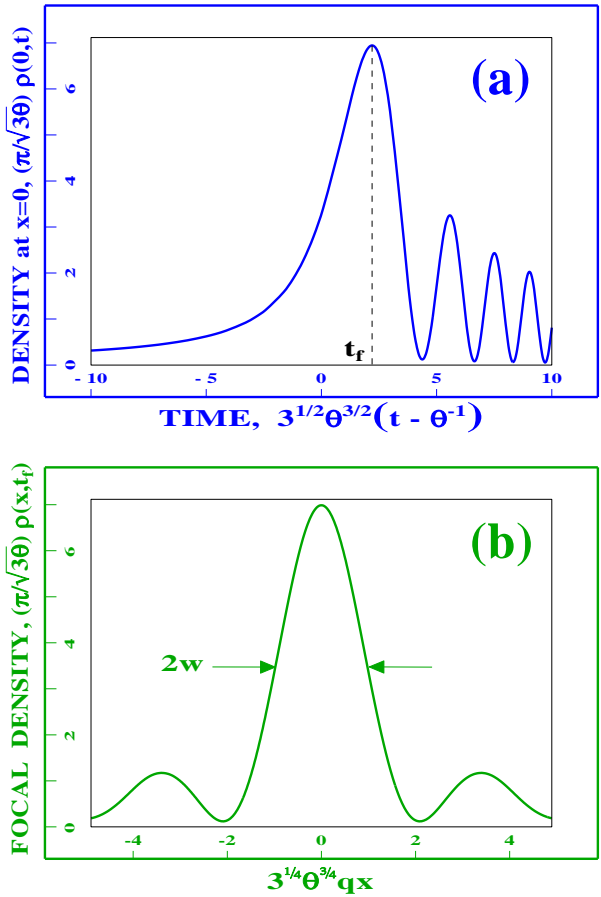}
\end{minipage}
\caption{Scaled asymptotic (as $\protect\theta \rightarrow \infty $) atom
densities in far-detuned focusing. (a) Time evolution of the atom density at 
$x=0$, $\protect\rho (0,t)$. (b) Density profile at the focal plane, $%
\protect\rho (x,t_{f})$. Time zero in the graph corresponds to the focal
point in the harmonic approximation.}
\label{fig01}
\end{figure}

In order to proceed, we need to determine the peak of the function $F\left(
0,\tilde{\omega}\right) /t$ from Eq. (\ref{a12a}) near $t\simeq \theta ^{-1}$
for $\theta ^{^{1}/_{2}}\gg 1$. First, times within the depth of focus $%
\left| t-\theta ^{-1}\right| $ are roughly determined by the requirement $%
\left| \tilde{\omega}\right| \sim 1$ or 
\begin{equation}
\left| t-\theta ^{-1}\right| \sim \theta ^{-\,^{3}/_{2}}  \label{a12c}
\end{equation}
and scale as $\sim $ $\theta ^{-\,^{1}/_{2}}$ times the focal plane
position. For $\theta ^{^{1}/_{2}}\gg 1,$ Eq. (\ref{a12c}) justifies the
inequality 
\begin{equation}
\left| t-\theta ^{-1}\right| \ll t\text{.}  \label{6p}
\end{equation}
Therefore, it is sufficient to replace the slowly-varying time dependence $%
t^{-1}$ by $\theta $ in Eqs. (\ref{a11c}) and (\ref{a12a}), taking the
lowest order limit of (\ref{a11d}) near $t\simeq \theta ^{-1}$. In other
words we replace Eqs. (\ref{a11c}) and (\ref{a11d}) by 
\begin{mathletters}
\label{6p}
\begin{eqnarray}
\tilde{x}_{a} &\sim &3^{^{1}/_{4}}\theta ^{^{3}/_{4}}x\text{ and }
\label{6p''} \\
\tilde{\omega}_{a} &\sim &3^{^{1}/_{2}}\theta ^{^{3}/_{2}}\left( \theta
^{-1}-t\right) ,  \label{6p'}
\end{eqnarray}
respectively, to arrive at the density near the focus, 
\end{mathletters}
\begin{equation}
\rho \left( x,t\right) \sim 3^{^{1}/_{2}}\pi ^{-1}\theta ^{^{1}/_{2}}F\left[ 
\tilde{x}_{a},\tilde{\omega}_{a}\right] \text{.}  \label{7}
\end{equation}
The subscript $a$ reminds us these results are asymptotic. Equation (\ref{7}%
) can be evaluated numerically. Unlike Eqs. (\ref{a12a}) and (\ref{a13a}),
Eq. (\ref{7}) is independent of an explicit dependence on $t$ and,
therefore, is a universal (scaled) density function. Inserting Eqs. (\ref
{a13}) and (\ref{6p'}) into (\ref{7}), one finds that the asymptotic time
evolution of the density along $x=0$ can be written analytically as 
\begin{eqnarray}
&&\left. \rho \left( 0,t\right) \sim 2^{-3\,}(3\theta )^{^{1}/_{2}}\pi
\left| \tilde{\omega}_{a}\right| \right.   \label{7'} \\
&&\times \left| J_{-\,^{1}/_{4}}\left( \tilde{\omega}_{a}^{2}/8\right)
+i^{^{3}/_{2}}sign\left( \tilde{\omega}_{a}\right) J_{^{1}/_{4}}\left( 
\tilde{\omega}_{a}^{2}/8\right) \right| ^{2}
\end{eqnarray}
and is shown in Fig. \ref{fig01}.

Approximate expressions for the focal parameters are derived by finding the
maximum of the function $F\left( 0,\tilde{\omega}_{a}\right) =|f\left( 0,%
\tilde{\omega}_{a}\right) |^{2}$ either numerically using Eq. (\ref{7'}) or
graphically from Fig. \ref{fig01}. We find that the maximum $F\left( 0,%
\tilde{\omega}_{f}\right) \approx 6.94$ occurs at $\tilde{\omega}_{f}\approx
-2.20$. Hence, the asymptotic focal time is given by inverting Eq. (\ref{6p'}%
), 
\begin{equation}
t_{f}\sim \theta ^{-1}\left[ 1-\tilde{\omega}_{f}\left( 3\theta \right)
^{-\,^{1}/_{2}}\right] \simeq \theta ^{-1}\left[ 1+1.27\theta ^{-\,^{1}/_{2}}%
\right] .  \label{a14}
\end{equation}
The asymptotic atom distribution at the focal plane, $\rho \left(
x,t_{f}\right) ,$ putting $\tilde{\omega}_{a}=\tilde{\omega}_{f}$ into Eq. (%
\ref{7}), is also plotted in Fig. \ref{fig01}. For the peak density we have 
\begin{equation}
\rho \left( 0,t_{f}\right) \sim 3^{^{1}/_{2}}\pi ^{-1}F\left( 0,\tilde{\omega%
}_{f}\right) \theta ^{^{1}/_{2}}\simeq 3.83\theta ^{^{1}/_{2}}.  \label{a15}
\end{equation}
The asymptotic focal spot size $w$ can be expressed through the half-width $%
\tilde{x}_{f}$ of the function $F\left( \tilde{x}_{a},\tilde{\omega}%
_{f}\right) ,$ which is found to be $\tilde{x}_{f}\approx 0.979$. Inverting
Eq. (\ref{6p''}) gives the expression 
\begin{equation}
w\sim 3^{-\,^{1}/_{4}}\tilde{x}_{f}\theta ^{-\,^{3}/_{4}}\simeq 0.744\theta
^{-\,^{3}/_{4}}.  \label{a16}
\end{equation}
The asymptotic depth of focus, $\Delta t$, defined by Eqs. (\ref{depth1})
and (\ref{depth2}), is evaluated using Eq. (\ref{7'}) by finding the two
values $\tilde{\omega}_{\pm }$ on either side of $\tilde{\omega}_{f}\approx
-2.20$ for which $F\left( 0,\tilde{\omega}_{\pm }\right) =F\left( 0,\tilde{%
\omega}_{f}\right) /2\approx 3.47$. These are $\tilde{\omega}_{+}\approx
-3.42$ and $\tilde{\omega}_{-}\approx -0.115,$ giving a depth of focus from
inverting Eq. (\ref{6p'}) of 
\begin{equation}
\Delta t\sim -3^{-\,^{1}/_{2}}(\tilde{\omega}_{+}-\tilde{\omega}%
_{-})\theta ^{-\,^{3}/_{2}}\simeq 1.91\theta ^{-\,^{3}/_{2}}\text{.}
\label{a151}
\end{equation}
Equations (\ref{6p})-(\ref{a151}) constitute the Method 3 results describing
the SW lens near $x=2\pi m$ and $t=\theta ^{-\,1}$.
In much the same way, we can use Methods 2 and 3 to derive expressions for
the background density, defined as the density $\rho (x_{b},t)$ at the
midpoints between periodic focuses, $x_{b}=$ $x_{m}+\pi $. We need to find
the wave function, Eq. (\ref{a7}), at the point $x_{b}=$ $\pi $. Since no
singularity arises in the integral for $t\lesssim $ $t_{f}$ when the
integrand is expanded around $x^{\prime }=\pi $, only the harmonic term is
needed to find the wave function, where $\cos \left( x^{\prime }\right)
\simeq (x^{\prime }-\pi )^{2}/2-1$. We find the approximate wave function, $%
\sim (1+t\theta )^{-1/2}$, and background density, 
\begin{equation}
\rho (\pi ,t)\sim (1+t\theta )^{-1}\text{,}  \label{back}
\end{equation}
correct to order $\theta ^{-2}$. This is the Method 2 result. This
expression allows us to find the asymptotic density contrast of importance
for lithography, $c\left( t\right) $. The contrast is defined\cite{61} to be
the ratio of the atomic density at the focal points $x_{m}$ to the background%
$,$ 
\begin{equation}
c\left( t\right) =\rho \left( 0,t\right) /\rho \left( \pi ,t\right) ,
\label{contrast}
\end{equation}
where $\rho \left( 0,t\right) $ is given by Eqs. (\ref{a13}) and (\ref{a13a}%
) for Method 2.

In addition, putting the asymptotic focal time (\ref{a14}) from Method 3
into Eq. (\ref{back}), the asymptotic background density at the focus is $%
\rho (\pi ,t_{f})\sim (1+t_{f}\theta )^{-1},$ or 
\begin{equation}
\rho (\pi ,t_{f})\sim 0.5(1-0.635\theta ^{-\,^{1}/_{2}})\text{.}  \label{102}
\end{equation}
>From Eq. (\ref{contrast}) the asymptotic contrast ratio at the focus is
(within the accuracy of this calculation) 
\begin{equation}
c(t_{f})=\frac{\rho \left( 0,t_{f}\right) }{\rho (\pi ,t_{f})}\sim
7.66\theta ^{^{1}/_{2}}\text{.}  \label{103}
\end{equation}
Equations (\ref{102}) and (\ref{103}) are a Method 3 result.
\begin{figure}[tb!]
\centering
\begin{minipage}{8.0cm}
\epsfxsize= 8 cm \epsfysize= 8.11 cm \epsfbox{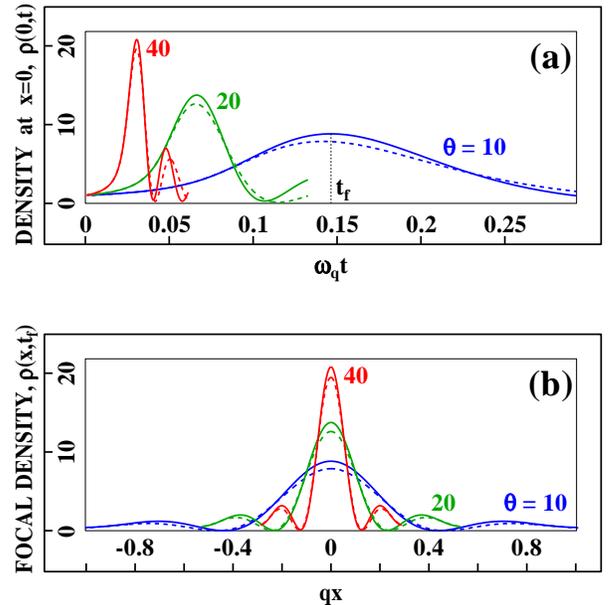}
\end{minipage}
\caption{Far-detuned focusing. Comparison between exact densities by the
Fourier method (Method 1, solid lines) and approximate asymptotic densities
near $x=0$ (Method 2, dashed lines) for different values of the pulse area $%
\protect\theta $. $(a)$ Time evolution of the atom density at $x=0$, $%
\protect\rho (0,t)$. $(b)$ Density profile at the focal plane, $\protect\rho %
(x,t_{f})$.}
\label{fig02}
\end{figure}

To summarize, we now have three methods to find the lens parameters, each
with different levels of numerical effort required to use it. The relevant
equations for these methods are listed in Table 1:

Table 1. Techniques and equations for three different methods of calculating
thin SW lens parameters. The numbers in the Table correspond to
equations in the text.
 
\[
\begin{tabular}{|c|c|c|c|}
\hline
Method & $
\begin{array}{c}
\text{Numerical} \\ 
\text{Sum}
\end{array}
$ & $
\begin{array}{c}
\text{Numerical} \\ 
\text{Integral}
\end{array}
$ & $
\begin{array}{c}
\text{Analytical} \\ 
\text{Expression}
\end{array}
$ \\ \hline
\multicolumn{1}{|l|}{1. Exact Fourier} & \ref{5b} &  &  \\ \hline
\multicolumn{1}{|l|}{2.$
\begin{array}{c}
\text{Asymptotic} \\ 
\left| x-m\pi \right| \ll 1 \\ 
t>0
\end{array}
$} &  & \ref{a12} & $
\begin{array}{c}
\text{\ref{a13a}} \\ 
\text{\ref{back}}
\end{array}
$ \\ \hline
\multicolumn{1}{|l|}{3.$
\begin{array}{c}
\text{Asymptotic} \\ 
\left| x-m\pi \right| \ll 1 \\ 
t\sim \theta ^{-\,1}
\end{array}
$} &  & \ref{7} & $
\begin{array}{c}
\text{\ref{7'}-\ref{a151}} \\ 
\text{\ref{102}-\ref{103}}
\end{array}
$ \\ \hline
\end{tabular}
\]
\begin{figure}[tb!]
\centering
\begin{minipage}{8.0cm}
\epsfxsize= 8 cm \epsfysize= 16 cm \epsfbox{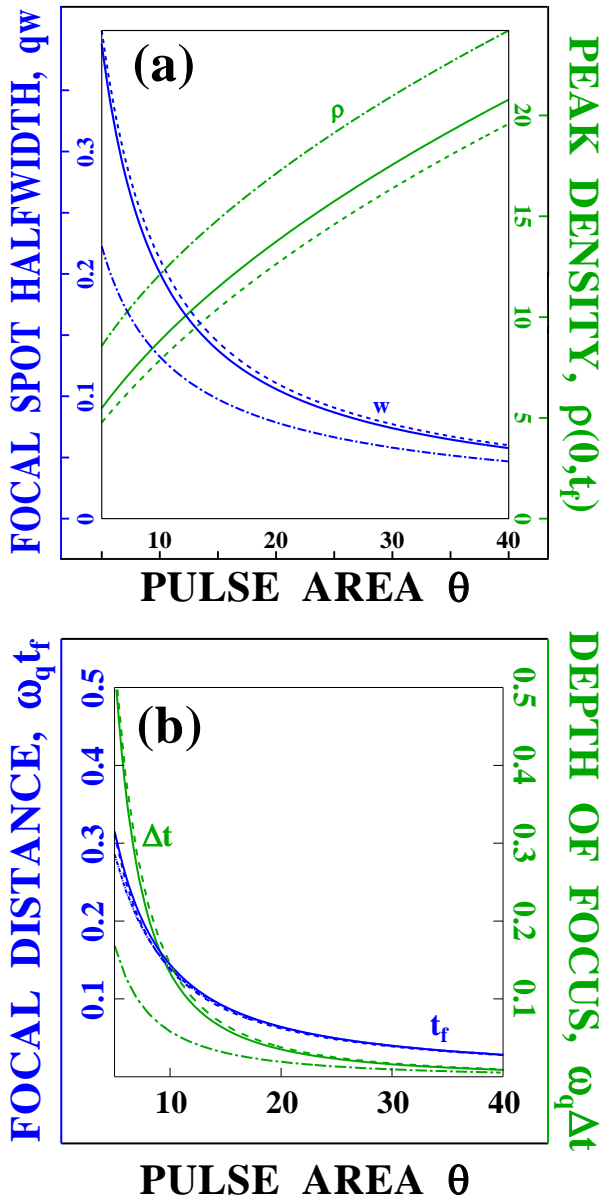}
\end{minipage}
\caption{Pulse area dependences of the lens parameters in far-detuned
focusing: $(a)$ Peak focal densities $\protect\rho (0,t_{f})$ (curves $%
\protect\rho $) and spot sizes $w$ (HWHM of the density profiles at the
focal planes). $(b)$ Focal distances $t_{f}$ and depths of focus $\Delta t$.
Computational methods: 1. Exact Fourier (solid), 2. Asymptotic diffraction
near $x=0$ (dashed), 3. Asymptotic diffraction near $x=0$ and $t=\protect%
\theta ^{-1}$ (dot-dashed). Note that the three methods give almost
identical result for $t_{f}$.}
\label{fig03}
\end{figure}

In Fig. \ref{fig02} the exact atom density at the center $\rho \left(
0,t\right) $ and atom distribution at the focal plane $\rho \left(
x,t_{f}\right) $, as calculated by Method 1, are compared with the
approximate expressions from Method 2. The convergence of the Method 2
result as $\theta $ increases is evident. In Figs. \ref{fig03} the pulse
area dependences of the focal parameters, as calculated by the three
methods, are compared. The accuracy of Method 2 is generally better than
that of Method 3, but all methods converge for $\theta ^{1/2}\gg 1$. These
results are discussed further in Sec. V, including the easily explained,
constant offsets of the asymptotic peak densities in Fig. \ref{fig03}a.

\section{Resonant focusing}

We now consider the density modulation and focusing caused by a resonant
standing wave field acting on a two-level atom. If the pulse duration
satisfies the inequality 
\begin{equation}
\tau \ll \min \{\left( \omega _{k}\theta \right) ^{-1},\left| \Delta \right|
^{-1},\Gamma ^{-1}\},  \label{9}
\end{equation}
[the field area $\theta $ for a resonant SW field is defined below by Eq. (%
\ref{7p})], then during the interaction the ground and excited state wave
functions, $\psi \left( x,t\right) $ and $\psi _{e}\left( x,t\right) $,
evolve in an interaction representation according to 
\begin{mathletters}
\label{10}
\begin{eqnarray}
\dot{\psi}\left( x,t\right) &=&-2i\chi g\left( t\right) \cos \left(
kx\right) \psi _{e}\left( x,t\right)  \label{10a} \\
\dot{\psi}_{e}\left( x,t\right) &=&-2i\chi g\left( t\right) \cos \left(
kx\right) \psi \left( x,t\right) .  \label{10b}
\end{eqnarray}
By satisfying Eq. (\ref{9}), we are assured that the resonant atom-field
interaction is in the Raman-Nath regime and that the pulse duration is
shorter than the excited state lifetime, avoiding the complications of
saturation and momentum space diffusion. If before the interaction the
incident wave function is uniform in the ground state, $\psi \left(
x,0^{-}\right) =1,$ then just after the interaction 
\end{mathletters}
\begin{equation}
\left( 
\begin{tabular}{l}
$\psi _{e}(x,0^{+})$ \\ 
$\psi (x,0^{+})$%
\end{tabular}
\right) =\left( 
\begin{tabular}{l}
$-i\sin [\left( \theta /2\right) \cos \left( kx\right) ]$ \\ 
$\cos [\left( \theta /2\right) \cos \left( kx\right) ]$%
\end{tabular}
\right) ,  \label{a2}
\end{equation}
where the effective pulse area for the resonant atom-field interaction is
defined as 
\begin{equation}
\theta =4\chi \int_{-\infty }^{\infty }dtg\left( t\right) .  \label{7p}
\end{equation}

The subsequent free space evolution and radiative decay of the system were
recently analyzed\cite{412} using the Fourier method for the closed
two-level scheme ({\it i.e.}, the excited state $\left| e\right\rangle $
decays only to $\left| g\right\rangle $)\cite{413}. However, the results
that are relevant to the problem discussed here pertain to any excited state
decay scheme. In the following paragraphs we summarize these results in
order to apply the three methods of Sec. II to the resonant case. Our goal
is to separate two terms in the free evolution of the ground state: (1) the
strict Hamiltonian evolution, resulting from the ground state amplitude
produced coherently by the SW pulse, and (2) the terms which result from
spontaneous decay of the excited state. Such a separation is straightforward
by the Fourier technique, Method 1. Once this is done, we can use a
diffraction analysis to obtain asymptotic results for the focal density
caused by the Hamiltonian term, $\psi (x,0^{+})$ in Eq. (\ref{a2}). The
diffraction analysis, Methods 2 and 3 from above, leads to analytic
expressions for the resonant SW lens parameters and, anticipating the
results, sheds light on the Rabi-like oscillation of the lens parameters as
a function of $\theta $.

After the excited state decays (in a time $t\gtrsim \Gamma ^{-1}$ following
the pulse), the total atomic density $\rho \left( x,t\right) $ involves only
the ground state and is a periodic function of $x$ having period $\lambda /2$%
. The atomic system can no longer be described simply by a wave function
since we must trace the full density matrix over the spontaneously emitted
photons to form the reduced density matrix $\rho \left( x,x^{\prime
},t\right) $. However, the part of the density matrix which was produced
coherently by the pulse can still be described by evolving the wave function 
$\psi (x,0^{+})$ in Eq. (\ref{a2}) into $\psi \left( x,t\right) $. This was
done in Ref.\cite{412}. The total density $\rho \left( x,t\right) $, which
is formally the diagonal component $\rho \left( x,x,t\right) $ of the
reduced density matrix, can be expanded in a set of Fourier components, 
\begin{mathletters}
\label{11}
\begin{eqnarray}
\rho \left( x,t\right)  &=&\sum_{n=-\infty }^{\infty }\left[ \rho
_{n}^{\left( S\right) }\left( t\right) +\rho _{n}^{\left( D\right) }\left(
t\right) \right] e^{inqx}  \nonumber \\
&=&\rho ^{\left( S\right) }\left( x,t\right) +\rho ^{\left( D\right) }\left(
x,t\right)   \label{11a} \\
\rho ^{\left( S\right) }\left( x,t\right)  &=&\sum_{n=-\infty }^{\infty
}\rho _{n}^{\left( S\right) }\left( t\right) e^{inqx}  \label{11b} \\
\rho ^{\left( D\right) }\left( x,t\right)  &=&\sum_{n=-\infty }^{\infty
}\rho _{n}^{\left( D\right) }\left( t\right) e^{inqx}\text{.}  \label{11c}
\end{eqnarray}
Each ground state Fourier component, $\rho _{n}\left( t\right) =\rho
_{n}^{\left( S\right) }\left( t\right) +\rho _{n}^{\left( D\right) }\left(
t\right) ,$ consists of two terms: a stimulated term, $\rho _{n}^{\left(
S\right) }\left( t\right) $, caused by the free evolution of $\psi (x,0^{+})$%
, and a spontaneous term, $\rho _{n}^{\left( D\right) }\left( t\right) $.

After Fourier expansion of $\psi (x,0^{+})$, the stimulated Fourier
components of the density are given by squaring 
\end{mathletters}
\begin{equation}
\psi \left( x,t\right) =\sum_{n=-\infty }^{\infty }(-1)^{n}J_{2n}\left(
\theta /2\right) e^{inqx-in^{2}\omega _{q}t}\text{.}  \label{13}
\end{equation}
to form\cite{412} 
\begin{eqnarray}
\rho _{n}^{\left( S\right) }\left( t\right)  &=&\,^{1}/_{2}\left\{ J_{2n}
\left[ \theta \sin \left( n\omega _{q}t/2\right) \right] \right.   \nonumber
\\
&&+\left. (-1)^{n}J_{2n}\left[ \theta \cos \left( n\omega _{q}t/2\right) %
\right] \right\} \text{.}  \label{12}
\end{eqnarray}
In particular, the average stimulated density ($n=0$) is 
\begin{equation}
\rho _{0}^{\left( S\right) }\left( t\right) =\,^{1}/_{2}\left[ 1+J_{0}\left(
\theta \right) \right] \text{.}  \label{12'}
\end{equation}
Each of the stimulated terms, $\rho _{n}^{\left( S\right) }\left( t\right) ,$
is periodic in time with the Talbot period, $T=2\pi /\omega _{q}.$

The $nth$ spontaneous term, $\rho _{n}^{\left( D\right) }\left( t\right) $,
results directly from the decay of the $nth$ Fourier component of the
excited state population. Since this decay is accompanied by an atomic
recoil, each harmonic's spontaneous term for $n\neq 0$ acquires an
additional Doppler phase which depends on the momentum $\hbar {\bf k}_{r}$
of the emitted photon, $n\hbar q{\bf \hat{x}\cdot k}_{r}t/M$. The resulting
inhomogeneous dephasing after integrating over spontaneous emission
directions leads to a decay of the spontaneous part of the $n\neq 0$ Fourier
components on a time scale of the order of $T,$%
\begin{equation}
\rho _{n\neq 0}^{\left( D\right) }\left( t\gg T\right) \longrightarrow 0.
\end{equation}
In the spatial domain this is equivalent to a length scale on the order of
the Talbot distance, $L_{T}$. While we could use the complete results of
Ref. \cite{412} to analyze focusing including this spontaneous term for
times $t\lesssim T$, we choose instead to simplify the problem by waiting
exactly $j$ Talbot periods $(t\simeq jT\gg T)$, where $j$ is a positive
integer greater than one. As a result, one can neglect $\rho _{n\neq
0}^{\left( D\right) }\left( t\right) $ in Eqs. (\ref{11}).

The $n=0$ spontaneous Fourier component does not decay. For $n=0$ it follows
from the Fourier expansion of $\psi _{e}(x,0^{+})$, for example, that 
\begin{equation}
\rho _{0}^{\left( D\right) }\left( t\right) =\,^{1}/_{2}\left[ 1-J_{0}\left(
\theta \right) \right] \text{.}  \label{spon}
\end{equation}
Thus, in accordance with the conservation of probability, the total average
density is one by summing Eqs. (\ref{12'}) and (\ref{spon}), 
\begin{equation}
\rho _{0}\left( t\right) =\rho _{0}^{\left( S\right) }\left( t\right) +\rho
_{0}^{\left( D\right) }\left( t\right) =1.
\end{equation}
Combining these results, the total density is 
\begin{equation}
\rho \left( x,t\gg T\right) =1+\sum_{n\neq 0}\rho _{n}^{\left( S\right)
}\left( t\right) e^{inqx}\text{.}  \label{dens}
\end{equation}
We can now use Eqs. (\ref{11}), (\ref{12}), (\ref{spon}), and (\ref{dens})
as exact, Method 1 expressions to compare to an approximate diffraction
theory.

To present a diffraction theory of resonant focusing, we consider only the
stimulated ground state wave function $\psi \left( x,t\right) $ evolving
from $\psi (x,jT+0^{+})=\psi (x,0^{+})$ [Eq. (\ref{a2})], remembering that
it is first necessary to add an integer number of Talbot periods, $jT$, to
reach an ''initial'' time when the modulated spontaneous contribution can be
neglected. Moreover, we must add on the additional background density from
the spontaneous term, Eq. (\ref{spon}), to the final result.

The wave function, $\psi \left( x,0^{+}\right) =\cos [\left( \theta
/2\right) \cos \left( kx\right) ],$ is amplitude modulated; however, $\psi
\left( x,0^{+}\right) $ superposes two terms, each of which is phase
modulated, 
\begin{mathletters}
\label{a3}
\begin{eqnarray}
\psi \left( x,0^{+}\right) &=&\psi _{+}\left( x,0^{+}\right) +\psi
_{-}\left( x,0^{+}\right) ,  \label{a3a} \\
\psi _{\pm }\left( x,0^{+}\right) &=&\,^{1}/_{2}\exp \left[ \pm i\left(
\theta /2\right) \cos \left( kx\right) \right] \text{.}  \label{a3b}
\end{eqnarray}
These\ states can be seen as evolving independently, leading to the
components of $\psi \left( x,t\right) $ we denote as $\psi _{\pm }\left( x%
{\bf ,}t\right) .$ Therefore, the problem of focusing after this specific
type of amplitude modulation can be mapped onto the problem of focusing by a
phase grating. For times $\omega _{k}t\simeq \theta ^{-1}$, $\theta >0$,
near the even antinodes, $kx=0,\,\pm 2\pi ,\,\ldots ,$ the component $\psi
_{+}\left( x,t\right) $ is responsible for focusing while the component $%
\psi _{-}\left( x,t\right) $ evolves smoothly as a background density. Near
the odd antinodes, $kx=\pm \pi ,\,\pm 3\pi ,\,\ldots ,$ the role of the
components $\psi _{\pm }\left( x,t\right) $ is reversed. The stimulated
density is then given by 
\end{mathletters}
\begin{equation}
\rho ^{\left( S\right) }\left( x,t\right) =\left| \psi _{+}\left( x,t\right)
+\psi _{-}\left( x,t\right) \right| ^{2}
\end{equation}
and the total density, adding on the spontaneous background $\rho
_{0}^{\left( D\right) }\left( t\right) $, by 
\begin{equation}
\rho \left( x,t\right) =\,^{1}/_{2}\left[ 1-J_{0}\left( \theta \right) %
\right] +\rho ^{\left( S\right) }\left( x,t\right) \text{.}  \label{tot}
\end{equation}

Consider the atom density near $x=0$ (the even antinodes) for the
development of the asymptotic theory. Since the period of the component wave
functions $\psi _{\pm }\left( x,t\right) $ is twice as large as the far
detuned case [compare Eqs. (\ref{a6}) and (\ref{a3b})], the dimensionless
coordinate and time become 
\begin{equation}
x\rightarrow kx,\qquad t\rightarrow \omega _{k}t,  \label{a17}
\end{equation}
noting that $q=2k$ and $\omega _{q}=4\omega _{k}$. We again assume that $%
\theta ^{^{1}/_{2}}\gg 1$. For the $\psi _{+}\left( x,t\right) $ component,
free evolution in the coordinate representation including the lowest-order
spherical aberration gives 
\begin{equation}
\psi _{+}\left( x,t\right) =\left( 3/\theta \right) ^{^{1}/_{4}}\left( 4\pi
it\right) ^{-\,^{1}/_{2}}\exp \left( i\theta /2\right) f\left( \tilde{x},%
\tilde{\omega}\right) ,  \label{a18}
\end{equation}
where the function $f$ and variables $\tilde{\omega}$ and $\tilde{x}$ are
given by Eqs. (\ref{a11b} - \ref{a11d}).

To calculate $\psi _{-}\left( x,t\right) $, it is sufficient to consider
only the harmonic part of the potential near $x=0,$ just as we did for the
background term (\ref{back}) in the phase modulation case. Replacing $\psi
\left( x^{\prime },t^{\prime }\right) $ in the Fresnel-Kirchhoff equation (%
\ref{a7}) by $\psi _{-}\left( x^{\prime },0^{+}\right) $ and using the
expansion $\cos \left( x^{\prime }\right) \approx 1-x^{\prime 2}/2,$
one finds 
\begin{mathletters}
\label{a19}
\begin{eqnarray}
\psi _{-}\left( x,t\right) &=&\,^{1}/_{2}\left( 1+\theta t\right)
^{-\,^{1}/_{2}}\exp \left( -i\theta /2+i\tilde{x}_{-}^{2}\right) ,
\label{a19a} \\
\tilde{x}_{-} &=&\left[ 4t\left( 1+\theta t\right) \right] ^{-\,^{1}/_{2}}x.
\label{a19b}
\end{eqnarray}
This contribution remains of order unity at the focus, i.e., it has relative
weight $\theta ^{-\,^{1}/_{4}}$ compared to Eq. (\ref{a18}). Since Eq. (\ref
{a18}) is of higher accuracy than Eq. (\ref{a19a}) (relative corrections to
Eq. (\ref{a18}) are of order $\theta ^{-\,^{1}/_{2}}$), it is valid to add
Eqs. (\ref{a18}) and (\ref{a19a}) to form the total wave function and
calculate the density to absolute order $\theta ^{^{1}/_{4}}$.

Interference between $\psi _{+}\left( x,t\right) $ and $\psi _{-}\left(
x,t\right) $ leads to a new effect, a Rabi-like oscillation of the focused
atom distribution. Squaring $\psi (x,t)=\psi _{+}\left( x,t\right) +\psi
_{-}\left( x,t\right) $ to form $\rho ^{\left( S\right) }\left( x,t\right) $%
, we find that the asymptotic stimulated atom density near $x=0$ is 
\end{mathletters}
\begin{mathletters}
\label{a20}
\begin{eqnarray}
\rho ^{\left( S\right) }\left( x,t\right)  &\sim &\left( 4\pi t\right)
^{-1}\left( 3/\theta \right) ^{^{1}/_{2}}\left\{ F\left( \tilde{x},\tilde{%
\omega}\right) \right.   \nonumber \\
&&+\theta ^{^{1}/_{4}}\left[ 2t/\left( 1+\theta t\right) \right] ^{^{1}/_{2}}
\nonumber \\
&&\times \left[ \cos \left( \theta -\tilde{x}_{-}^{2}\right) f_{c}\left( 
\tilde{x},\tilde{\omega}\right) \right.   \nonumber \\
&&-\left. \left. \sin \left( \theta -\tilde{x}_{-}^{2}\right) f_{s}\left( 
\tilde{x},\tilde{\omega}\right) \right] \right\} ,  \label{a20a} \\
f_{c}\left( \tilde{x},\tilde{\omega}\right) +if_{s}\left( \tilde{x},\tilde{%
\omega}\right)  &\equiv &\left( 2\pi /i\right)
^{^{1}/_{2}}3^{-\,^{1}/_{4}}f\left( \tilde{x},\tilde{\omega}\right) .
\label{a20b}
\end{eqnarray}
Due to the symmetry between $\psi _{+}$ and $\psi _{-}$, this is the density
near the focus at both the even and odd antinodes, giving a total density of
spatial period $2\pi /q=\lambda /2$. Equations (\ref{a20}) are the Method 2
result for the resonant lens. After accounting for the relevant coordinate
and time scales (\ref{a17}), the dominant term, $|\psi _{+}|^{2}$, looks
exactly like the far detuned result of Sec. II and leads to the same lens
parameters if taken alone. The interference term has a relative amplitude $%
\theta ^{-\,^{1}/_{4}}$ near the focus ($t\simeq \theta ^{-1}$) when
compared to the $|\psi _{+}|^{2}$ term and oscillates sinusoidally with $%
\theta .$

Now we employ Method 3, the limit of this asymptotic result for $t\simeq
\theta ^{-1}$. Considering the interference term as a small correction to $%
|\psi _{+}|^{2}$, we apply the far detuned results [Eqs. (\ref{7}) and (\ref
{7'}), $\tilde{\omega}_{f}\approx -2.20$, and $\tilde{x}_{f}\approx 0.979$]
and Taylor expand Eqs. (\ref{a20}) around $\tilde{\omega}_{f}$ and $\tilde{x}%
_{f}$. The following asymptotic expressions can be derived for the focal
parameters: 
\end{mathletters}
\begin{mathletters}
\label{a21}
\begin{eqnarray}
t_{f} &\sim &\theta ^{-1}\left\{ 1-\tilde{\omega}\left( 3\theta \right)
^{-\,^{1}/_{2}}+3^{-\,^{1}/_{2}}\theta ^{-\,^{3}/_{4}}\right.   \nonumber \\
&&\times \left[ \partial ^{2}F\left( 0,\tilde{\omega}\right) /\partial 
\tilde{\omega}^{2}\right] ^{-1}\left[ \cos \left( \theta \right) \right.  
\nonumber \\
&&\times \partial f_{c}\left( 0,\tilde{\omega}\right) /\partial \tilde{\omega%
}  \nonumber \\
&&-\left. \left. \sin \left( \theta \right) \partial f_{s}\left( 0,\tilde{%
\omega}\right) /\partial \tilde{\omega}\right] \right\} _{\tilde{\omega}=%
\tilde{\omega}_{f}}  \nonumber \\
&\simeq &\theta ^{-1}\left\{ 1+1.27\theta ^{-\,^{1}/_{2}}\right. +\theta
^{-\,^{3}/_{4}}  \nonumber \\
&&\times \left[ -0.465\cos \left( \theta \right) \right.   \nonumber \\
&&+\left. \left. 0.110\sin \left( \theta \right) \right] \right\} ,
\label{a21a} \\
\rho ^{\left( S\right) }\left( 0,t_{f}\right)  &\sim &3^{^{1}/_{2}}\left(
4\pi \right) ^{-1}\theta ^{^{1}/_{2}}\left\{ F\left( 0,\tilde{\omega}%
_{f}\right) \right.   \nonumber \\
&&+\theta ^{-\,^{1}/_{4}}\left\{ \left[ \cos \left( \theta \right)
f_{c}\left( 0,\tilde{\omega}_{f}\right) \right. \right.   \nonumber \\
&&-\left. \left. \sin \left( \theta \right) f_{s}\left( 0,\tilde{\omega}%
_{f}\right) \right] \right\}   \nonumber \\
&\simeq &0.957\theta ^{^{1}/_{2}}\left\{ 1+\theta ^{-\,^{1}/_{4}}\left[
0.166\cos \left( \theta \right) \right. \right.   \nonumber \\
&&+\left. \left. 0.703\sin \left( \theta \right) \right] \right\} ,
\label{a21b} \\
w &\sim &3^{-\,^{1}/_{4}}\theta ^{-\,^{3}/_{4}}\left\{ \tilde{x}\right.  
\nonumber \\
&&+\theta ^{-\,^{1}/_{4}}\left[ \partial F\left( \tilde{x},\tilde{\omega}%
\right) /\partial \tilde{x}\right] ^{-1}  \nonumber \\
&&\times \left[ \cos \left( \theta \right) \left( ^{1}/_{2}f_{c}\left( 0,%
\tilde{\omega}\right) -f_{c}\left( \tilde{x},\tilde{\omega}\right) \right.
\right.   \nonumber \\
&&+\left( \partial f_{c}\left( 0,\tilde{\omega}\right) /\partial \tilde{%
\omega}\right) \left( \partial F\left( \tilde{x},\tilde{\omega}\right)
/\partial \tilde{\omega}\right)   \nonumber \\
&&\times \left. \left( \partial ^{2}F\left( 0,\tilde{\omega}\right)
/\partial \tilde{x}^{2}\right) ^{-1}\right)   \nonumber \\
&&-\left( \text{the same with replacements }\right.   \nonumber \\
&&\left. \left. \left. \cos \left( \theta \right) \left. \rightarrow \right.
\sin \left( \theta \right) ,\,\,f_{c}\left. \rightarrow \right. f_{s}\right) 
\right] \right\} _{\tilde{\omega}=\tilde{\omega}_{f},\tilde{x}=\tilde{x}_{f}}
\nonumber \\
&\simeq &0.744\theta ^{-\,^{3}/_{4}}\left\{ 1+\theta ^{-\,^{1}/_{4}}\left[
0.198\cos \left( \theta \right) \right. \right.   \nonumber \\
&&+\left. \left. 0.162\sin \left( \theta \right) \right] \right\} .
\label{a21c}
\end{eqnarray}
\begin{figure}[tb!]
\centering
\begin{minipage}{8.0cm}
\epsfxsize= 8 cm \epsfysize= 9.77 cm \epsfbox{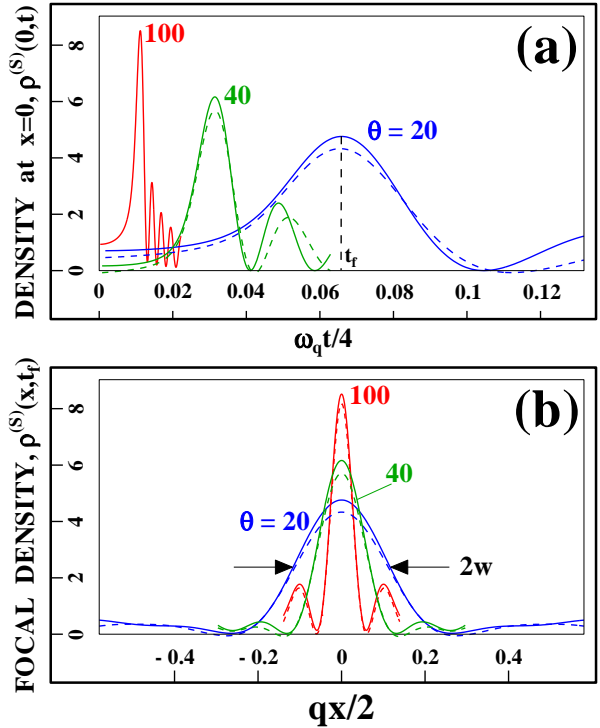}
\end{minipage}
\caption{Resonant focusing. Comparison between exact densities by the
Fourier method (Method 1, solid lines) and approximate asymptotic densities
near $x=0$ (Method 2, dashed lines) for different values of the pulse area $%
\protect\theta $. $(a)$ Time evolution of the atom density at $x=0$, $%
\protect\rho ^{(S)} (0,t)$. $(b)$ Density profile at the focal plane, $%
\protect\rho ^{(S)} (x,t_{f})$. The atom density is plotted excluding the
constant background term caused by spontaneous emission, $\protect\rho %
^{(D)}_{0} (t)=(1-J_{0}(\protect\theta ))/2$. }
\label{fig04}
\end{figure}

In Fig. \ref{fig04} the exact time dependence of the peak stimulated
density, $\rho ^{\left( S\right) }\left( x,t\right) $, and the stimulated
density profile at the focal plane, $\rho ^{\left( S\right) }\left(
x,t_{f}\right) $, as calculated numerically by Method 1 using Eqs. (\ref{11b}%
) and (\ref{12}) for several pulse areas, are compared with the approximate
diffraction expression of Method 2, Eq. (\ref{a20a}). A comparison of the
focal parameters as functions of the pulse area for the three different
methods is plotted in Fig. \ref{fig05}, clearly showing the oscillation of
the exact and approximate results with $\theta $. While a diffraction theory
may be unnecessary since an exact result can be calculated by the Fourier
method, the physical origin of the Rabi-like term, which is masked by the
Fourier result, has been revealed by the diffraction theory to be the
interference of the $\psi _{+}$ and $\psi _{-}$ components of the wave
function. Further discussion is again reserved for Sec. V.
\begin{figure}[tb!]
\centering
\begin{minipage}{8.0cm}
\epsfxsize= 8 cm \epsfysize= 9.77 cm \epsfbox{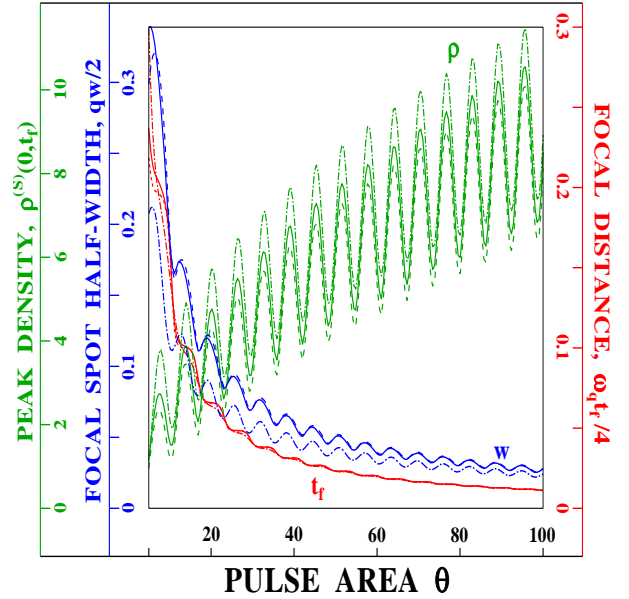}
\end{minipage}
\caption{Resonant focusing. Pulse area dependences of the focusing
parameters: Peak focal densities $[\protect\rho (0,t_{f})$ (curves $\protect%
\rho $)], spot sizes $w$ (HWHM of the density profiles at the focal planes),
and focal distances $t_{f}$. Computational methods: 1. Exact Fourier
(solid), 2. Asymptotic diffraction near $x=0$ (dashed), 3. Asymptotic
diffraction near $x=0$ and $t=\protect\theta ^{-1}$ (dot-dashed). The atom
density is plotted excluding the constant background term caused by
spontaneous emission, $\protect\rho ^{(D)}_{0} (t)=(1-J_{0}(\protect\theta
))/2$. }
\label{fig05}
\end{figure}

\section{Chromatic aberration and transverse velocity distributions}

\subsection{Chromatic aberration}

Chromatic aberration, the dispersion or wave length dependence of a lens'
properties, results from a finite distribution of longitudinal velocities (%
{\it i.e.}, de Broglie wave lengths) in an atom beam. In trap experiments
this type of aberration is avoided. However, for beam experiments detected
in the laboratory frame, the atomic distribution would be probed or
deposited at a certain distance from the interaction region, $L=Ut$. As a
result, the total atom density at $L$ will be an average of each velocity's
density at $L$ over the flux of atoms with that velocity. In general, to
calculate the density, we must know the longitudinal velocity distribution
or at least its statistical properties. Previous theoretical and
experimental publications that have discussed chromatic aberration in the
context of atom optical lenses have used heuristic, numerical, or Monte
Carlo simulation approaches \cite{6p,61,7,81,82,83b,84}. We attack the
problem from a different perspective, based on averaging the exact Fourier
components over the atomic flux distribution. The technique is easy to
apply, the results are simple to understand, and the conclusions have a
physical interpretation.

For the remainder of the paper, the {\it ideal} case will refer to the
monovelocity atomic beam with longitudinal velocity $U$ and infinitely
narrow longitudinal and transverse velocity distributions. In the previous
sections we have presented a full characterization of the thin SW lens
acting on the ideal beam. Two types of beams with more realistic
longitudinal distributions are typical and will be considered here, (1) a
velocity narrowed beam centered around some average velocity $U_{0}$ or (2)
a thermal beam with average speed $U_{0}$. In this section we consider only
focusing by the far-detuned standing wave. For this type of lens, the exact
expression for the density, Eq. (\ref{5b}), can be averaged over the proper
flux distribution function for a specified $L$ by numerical integration,
provided the paraxial approximation still holds: $\lambda _{dB,0}=2\pi \hbar
/(MU_{0})\ll \lambda $. In addition, the narrow velocity distribution case
allows for an approximate analytical solution to compare to the numerically
integrated result: by expanding the lens parameters around the average
values of the flux distribution, we can account for chromatic aberration
analytically using the density of the Fourier method.

In the laboratory frame the atoms interact with a SW field with a fixed
width $\sigma _{z}$ along the $z$ direction and are then detected at a fixed
distance $L$ from the lens. As a result, the pulse area $\theta $ and the
time of free flight $t$ after the lens depend on the atomic longitudinal
velocity $U$ as $U^{-1}$. The atomic density is defined as an average over
the single-particle atomic flux distribution $W\left( U\right) $. This
distribution is a measure of the probability, $W\left( U\right) dU,$ to find
an atom with velocity between $U$ and $U+dU$ passing through a plane
perpendicular to the $z$ axis. (One would divide $W\left( U\right) $ by the
atomic spatial density to get a true flux distribution in inverse units of
particles per unit time per unit area.) To derive quantitative results, we
need to specify $W\left( U\right) $ for a given experiment.

In this section we denote flux averages $\left\langle {}\right\rangle $ by 
\end{mathletters}
\begin{equation}
\left\langle F\left( U\right) \right\rangle =\int_{0}^{\infty }dUF\left(
U\right) W\left( U\right) ,  \label{17}
\end{equation}
taking into account that only atoms having positive velocity, $U>0$, have to
be included. Using this definition, the average velocity $U_{0}$, the
(dimensionless) relative velocity $u$, and the (dimensionless) flux width $v$
are given, respectively, as 
\begin{mathletters}
\label{16}
\begin{eqnarray}
U_{0} &=&\left\langle U\right\rangle ,  \label{16a} \\
u &=&\left( U-U_{0}\right) /U_{0},  \label{16b} \\
v &=&\left[ 2\left\langle u^{2}\right\rangle \right] ^{\,^{1}/_{2}}\text{.}
\label{16c}
\end{eqnarray}
The pulse area $\theta $ and the time of flight $t$ for each velocity class $%
U$ can be defined by reference to the average velocity $U_{0}$, 
\end{mathletters}
\begin{equation}
\theta =\theta _{0}U_{0}/U=\theta _{0}/(1+u)\text{ and }%
t=t_{0}U_{0}/U=t_{0}/(1+u)\text{,}  \label{162}
\end{equation}
where $\theta _{0}\simeq (-4\left| \chi \right| ^{2}/\Delta )(\sigma
_{z}/U_{0})$ and $t_{0}=L/U_{0}\ $would be computed using $U_{0}$. The
exact, flux-averaged density $\bar{\rho}\left( x,t_{0}\right) $ follows
immediately from Eq. (\ref{5b}), 
\begin{eqnarray}
\bar{\rho}\left( x,t_{0}\right)  &=&\left\langle \rho \left( x,t\right)
\right\rangle   \nonumber \\
&=&\sum_{n=-\infty }^{\infty }e^{inx}\int_{0}^{\infty }dUW\left( U\right)  
\nonumber \\
&&\times J_{n}\left[ \left( \theta _{0}U_{0}/U\right) \sin \left(
nt_{0}U_{0}/U\right) \right] \text{.}  \label{therm}
\end{eqnarray}
The integrals can be evaluated numerically, term-by-term, giving the
flux-averaged Fourier components of the density. If the distribution $%
W\left( U\right) $ has a non-zero width $v$, then the focusing by the SW
lens is degraded. We show this graphically below.

We start with the narrow velocity distribution case, 
\begin{equation}
v\ll 1\text{,}  \label{161}
\end{equation}
and derive approximate, analytical expressions from Eq. (\ref{therm}) to
compare to an exact numerical integration. Our goal is to determine the
effect of a flux width $v$ on the focus when compared to the ideal beam
results ($v\rightarrow 0$). Since $v\ll 1,$ we can expand the field area and
time, Eq. (\ref{162}), as 
\begin{equation}
\theta \approx \theta _{0}\left( 1-u+u^{2}\right) ,\,\,t\approx t_{0}\left(
1-u+u^{2}\right) .  \label{19}
\end{equation}
Substituting values (\ref{19}) into $\rho \left( x,t\right) $ of Eq. (\ref
{5b}), expanding in powers of $u,$ and applying Eqs. (\ref{therm}) and (\ref
{16}), one finds to order $v^{2}$ that 
\begin{mathletters}
\label{20}
\begin{eqnarray}
\bar{\rho}\left( x,t_{0}\right)  &\simeq &\rho \left( x,t_{0}\right)
+v^{2}\rho _{1}\left( x,t_{0}\right) ,  \label{20a} \\
\rho _{1}\left( x,t\right)  &=&\,^{1}/_{4}\sum_{n=-\infty }^{\infty }\left\{
\theta _{0}\left[ 4nt\cos \left( nt\right) +\left( 2-\left( nt\right)
^{2}\right) \right. \right.   \nonumber \\
&&\times \left. \sin \left( nt\right) \right] J_{n}^{\prime }\left[ \theta
_{0}\sin \left( nt\right) \right] +\theta _{0}^{2}\left[ \sin \left(
nt\right) \right.   \nonumber \\
&&+\left. \left. nt\cos \left( nt\right) \right] ^{2}J_{n}^{\prime \prime }%
\left[ \theta _{0}\sin \left( nt\right) \right] \right\} e^{inx},
\label{20b}
\end{eqnarray}
where $\rho _{1}\left( x,t_{0}\right) $ is the lowest-order correction to
the density at $L$ as a result of the flux distribution.

To order $v^{2},$ four of the focal parameters (the focal distance $t_{f},$
defined as the first maximum of the function $\bar{\rho}\left(
0,t_{0}\right) ,$ the peak density $\bar{\rho}(0,t_{f}),$ the spot size $w,$
defined as the lowest root of the equation $\bar{\rho}(w,t_{f})=\bar{\rho}%
(0,t_{f})/2$, and the depth of focus $\Delta t,$ defined by $\bar{\rho}%
(0,t_{\pm })=\bar{\rho}(0,t_{f})/2$), are given by 
\end{mathletters}
\begin{mathletters}
\label{21}
\begin{eqnarray}
t_{f} &\simeq &t_{0,f}+v^{2}t_{1},  \label{21a} \\
t_{1} &=&-\left[ \frac{\partial \rho _{1}\left( 0,t\right) /\partial t}{%
\partial ^{2}\rho \left( 0,t\right) /\partial t^{2}}\right] _{t=t_{0,f}};
\label{21b} \\
\bar{\rho}\left( 0,t_{f}\right)  &\simeq &\rho \left( 0,t_{0,f}\right)
+v^{2}\rho _{1}\left( 0,t_{0,f}\right) ;  \label{21c} \\
w &\simeq &w_{0}+v^{2}w_{1},  \label{21d} \\
w_{1} &=&\left\{ \left[ \partial \rho \left( w,t\right) /\partial w\right]
^{-1}\left[ ^{1}/_{2}\rho _{1}\left( 0,t\right) \right. \right.   \nonumber
\\
&&-\rho _{1}\left( w,t\right)   \nonumber \\
&&-\left. \left. t_{1}\partial \rho \left( w,t\right) /\partial t\right]
\right\} _{t=t_{0,f}.w=w_{0}},  \label{21e} \\
\Delta t &=&\Delta t_{0}+v^{2}\Delta t_{1},  \label{21f} \\
\Delta t_{1} &=&\frac{\,^{1}/_{2}\rho _{1}\left( 0,t_{0,f}\right) -\rho
_{1}\left( 0,t_{0,+}\right) }{\left[ \partial \rho \left( 0,t\right)
/\partial t\right] _{t=t_{0,+}}}  \nonumber \\
&&-\frac{\,^{1}/_{2}\rho _{1}\left( 0,t_{0,f}\right) -\rho _{1}\left(
0,t_{0,-}\right) }{\left[ \partial \rho \left( 0,t\right) /\partial t\right]
_{t=t_{0,-}}},  \label{21g}
\end{eqnarray}
where $t_{0,f}$, $w_{0},$ and $\Delta t_{0}=t_{0,+}-t_{0,-}$ are the focal
distance, focal spot size, and the depth of focus, respectively, for $\theta
_{0}$ as calculated in Sec. II for the monovelocity beam. It is significant
that Eqs. (\ref{20}) and (\ref{21}) depend only on the dimensionless flux
width $v$ and not on the exact form of the distribution $W(U)$. The pulse
area dependences of the correction coefficients, $t_{1},\,\rho _{1}\left(
0,t_{0,f}\right) ,\,$and $w_{1}$, are shown in Fig. \ref{fig06}. To verify
Eqs. (\ref{20}) and (\ref{21}) for small $v$, we can now quantitatively
compare the approximate expressions to the exact, numerically averaged
density, Eq. (\ref{therm}), for a physically reasonable and mathematically
convenient distribution function.
\begin{figure}[tb!]
\centering
\begin{minipage}{8.0cm}
\epsfxsize= 8 cm \epsfysize= 7.31 cm \epsfbox{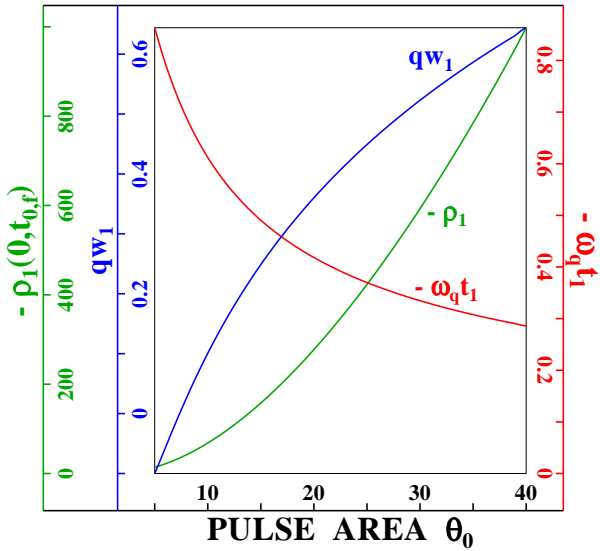}
\end{minipage}
\caption{Chromatic aberration for focusing atomic beams. Correction
coefficients, given in Eqs. (\ref{21}), as functions of the pulse area $%
\protect\theta _{0}$, assuming a narrow longitudinal flux distribution, $%
v\ll 1$: Peak focal density $\protect\rho _{1}(t_{0,f})$, focal distance $%
t_{1}$, and spot size $w_{1}$ corrections. These coefficients multiply $%
v^{2} $ to form the lowest-order corrections to the lens parameters. }
\label{fig06}
\end{figure}

For example, we can find the flux distribution $W_{lM}(U)$ corresponding to
the local Maxwellian velocity distribution, 
\end{mathletters}
\begin{equation}
W_{lM^{\prime }}\left( U\right) =\frac{1}{\sqrt{\pi }V}\exp \left[ -\left( U-%
\widetilde{U}_{0}\right) ^{2}/V^{2}\right] .  \label{171}
\end{equation}
The flux distribution, average velocity, and (dimensionless) flux width that
follow from $W_{lM^{\prime }}\left( U\right) $ can be written with accuracy $%
\exp \left( -\widetilde{U}_{0}^{2}/V^{2}\right) $ as 
\begin{mathletters}
\label{172}
\begin{eqnarray}
W_{lM}\left( U\right) &=&\frac{U}{\sqrt{\pi }\widetilde{U}_{0}V}\exp \left[
-\left( U-\widetilde{U}_{0}\right) ^{2}/V^{2}\right] ,  \label{172a} \\
U_{0} &=&\widetilde{U}_{0}\left( 1+V^{2}/2\widetilde{U}_{0}^{2}\right) ,
\label{172b} \\
v &=&\frac{V}{\widetilde{U}_{0}}\frac{\sqrt{1-V^{2}/2\widetilde{U}_{0}^{2}}}{%
1+V^{2}/2\widetilde{U}_{0}^{2}}.  \label{172c}
\end{eqnarray}
As long as $V/\widetilde{U}_{0}\ll 1,$ we can be assured that $v\ll 1$.
Distribution (\ref{172a}) can be inserted into Eq. (\ref{therm}) for $\bar{%
\rho}\left( x,t_{0}\right) $, and the exact density can be evaluated at the
time $t_{0,f}$ for different values of the distribution width $v$ using the
parametrization of Eq. (\ref{172c}). Alternatively, the precise,
flux-averaged focal time, $t_{f}=L_{f}/U_{0}\neq t_{0,f}$, can be found
numerically from the first maximum of $\bar{\rho}\left( 0,t_{0}\right) $ in
time and used to evaluate the density and lens parameters. The peak
reduction and spot size increase for two pulses areas at $t_{0,f}$ are
plotted as a function of $v$ in Fig. \ref{fig07}. In Fig. \ref{fig08} the
density profile at the true focus, $\bar{\rho}\left( x,t_{f}\right) $, for
one of these pulse areas is shown with and without chromatic aberration.
Both figures are explained in detail in Sec. V.
\begin{figure}[tb!]
\centering
\begin{minipage}{8.0cm}
\epsfxsize= 8 cm \epsfysize= 6.44 cm \epsfbox{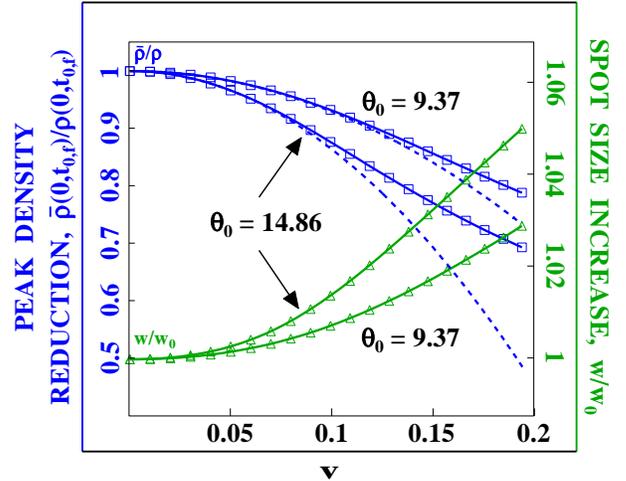}
\end{minipage}
\caption{Narrow longitudinal flux distribution in atomic focusing. Exact
reduction in the peak focal density, $\bar{\protect\rho}(0,t_{0,f})/\protect%
\rho (0,t_{0,f})$ (boxes), and exact increase in the spot size, $w/w_{0}$
(triangles), as functions of the longitudinal flux width $v$ for the flux
distribution $W_{lM}(U)$ and for two pulse areas, $\protect\theta _{0}=9.37$
and $14.86$. These pulse areas would produce spot sizes of $10$ $nm$ and $%
6.5 $ $nm$, respectively, in Na for $v=0$. The dashed curve for $\bar{%
\protect\rho}(0,t_{0,f})/\protect\rho (0,t_{0,f})$ is the approximate,
analytical dependence for small $v$ from Eq. (\ref{21c}) or Fig. \ref{fig06}%
. }
\label{fig07}
\end{figure}
\begin{figure}[tb!]
\centering
\begin{minipage}{8.0cm}
\epsfxsize= 8 cm \epsfysize= 5.5 cm \epsfbox{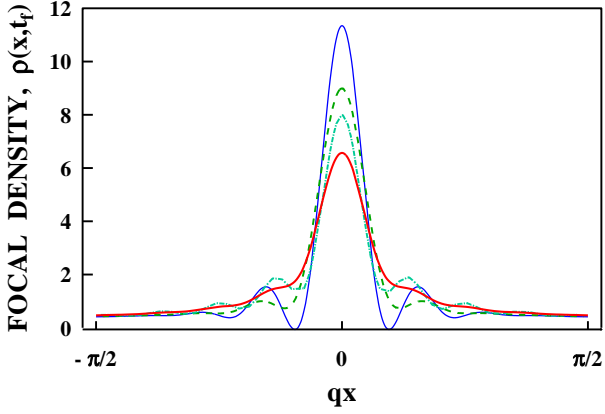}
\end{minipage}
\caption{Exact focal density profiles, $\protect\rho (x,t_{f})$, in the
range $|qx|\leq \protect\pi /2$ for $\protect\theta _{0}=14.86$ with and/or
without the effects of both chromatic aberration and angular divergence. The
ideal atomic beam - $v=0$, $V_{x}=0$ - gives a spot size of $qw=0.139$, or $%
6.5$ $nm$ in sodium (thin solid line). Pure chromatic aberration, $W_{lM}(U)$
longitudinal flux distribution - $v=0.194$ and $V_{2}=0$ (dotted line). Pure
angular divergence, $P_{2}$ (Maxwellian) transverse velocity distribution - $%
v=0$ and $V_{2}=1$ (dashed line). Combined chromatic aberration and angular
divergence - $v=0.194$ and $V_{2}=1$ (thick solid line). The transverse
width, $V_{2}=1$, corresponds to the single-photon recoil limit. See Table 2
for other quantitative details.}
\label{fig08}
\end{figure}

Next, we consider the other case of experimental interest, the Maxwellian
thermal beam having a flux distribution 
\end{mathletters}
\begin{equation}
W_{M}\left( U\right) =2U^{3}V^{-4}\exp \left( -U^{2}/V^{2}\right) ,
\label{22}
\end{equation}
where $V$ is now a thermal velocity. For a thermal flux, the distribution
width $v$, still defined by Eq. (\ref{16c}), is not small, 
\begin{equation}
v=\sqrt{64/9\pi -2}\approx 0.513.  \label{221}
\end{equation}
On the other hand, since $v\lesssim 1$, we expect that the order of
magnitude of the density's peak amplitude, the spot size (or spatial
resolution) of the lens, and the focal position should be the same. In
analogy with the analysis above, we consider the atomic density at a
distance $L$ from the SW lens, where $L=Vt_{0}$ is now defined by the free
evolution time $t_{0}$ for an atom with thermal velocity $V$. (The average
thermal flux velocity $U_{0}$ is given by $3\sqrt{\pi }V/4\approx 1.33V.$)

\begin{figure}[tb!]
\centering
\begin{minipage}{8.0cm}
\epsfxsize= 8 cm \epsfysize= 8 cm \epsfbox{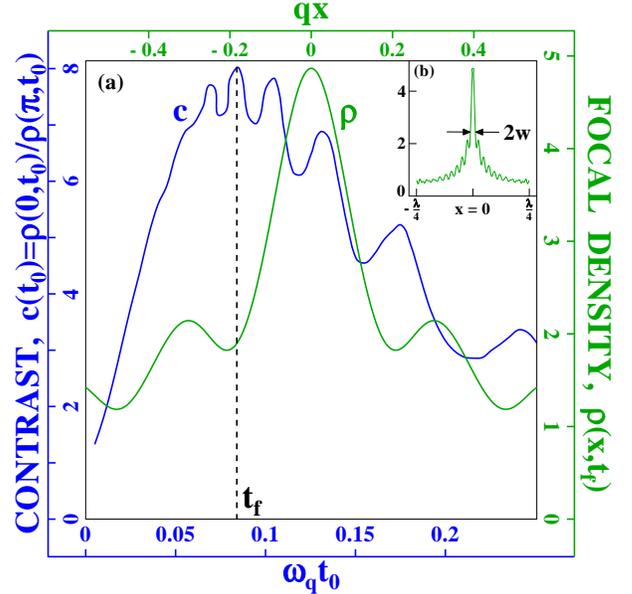}
\end{minipage}
\caption{Thermal longitudinal beam focusing. (a) Time evolution of the
atomic density contrast, $c(t_{0})=\bar{\protect\rho }(0,t_{0})/ \bar{%
\protect\rho }(\protect\pi ,t_{0})$, and the density profile at the focal
plane, $\bar{\protect\rho }(0,t_{f})$. (b) Atomic density profile at the
focus, $\bar{\protect\rho }(x,t_{f})$, plotted in the range $-\protect\lambda
/4\leq x\leq \protect\lambda /4.$ The pulse area for these curves, $\protect%
\theta _{0} =23.0$, is needed to produce a spot size of $w=6.5$ $nm$ in Na
using a thermal beam. }
\label{fig09}
\end{figure}

\begin{figure}[tb!]
\centering
\begin{minipage}{8.0cm}
\epsfxsize= 8 cm \epsfysize= 7.23 cm \epsfbox{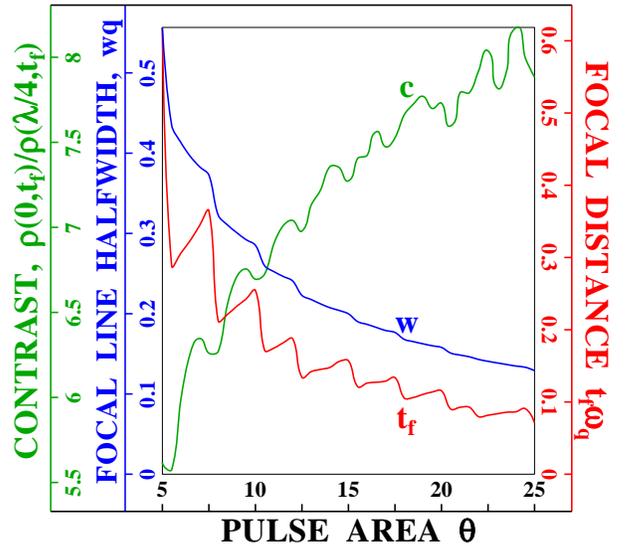}
\end{minipage}
\caption{Thermal longitudinal beam focusing. Exact pulse area dependence of
the focusing parameters: Focal density contrast $c(t_{f})=\bar{\protect\rho }%
(0,t_{f}) /\bar{\protect\rho }(\protect\pi ,t_{f})$, focal distance $t_{f}$,
and spot size $w$ (HWHM of the density profile at the focal plane).}
\label{fig10}
\end{figure}

We can now use the thermal flux distribution (\ref{22}) in Eq. (\ref{therm})
to calculate the various, exact lens parameters. To find the focus for this
distribution, instead of monitoring the peak atomic density $\bar{\rho}%
\left( 0,t_{0}\right) $, we monitor the thermally-averaged density contrast $%
\bar{c}\left( t_{0}\right) =\bar{\rho}\left( 0,t_{0}\right) /\bar{\rho}%
\left( \pi ,t_{0}\right) $ from Eq (\ref{contrast}). An example of $\bar{c}%
\left( t_{0}\right) $ for $\theta _{0}=23.0$ is shown in Fig. \ref{fig09}.
The focal plane position is defined by the time when $\bar{c}\left(
t_{0}\right) $ is largest, $t_{0}=t_{f}$ ($L=Vt_{f}$). The atomic density at 
$t_{f}$, $\bar{\rho}\left( x,t_{f}\right) $, for this pulse area is also
shown in Fig. \ref{fig09}. Several lens parameters as functions of $\theta
_{0}$ are shown in Fig. \ref{fig10}, including the focal time $t_{f}$, the
contrast at the focal plane $c\left( t_{f}\right) $, and the spot size $w$.

\subsection{Transverse velocity distributions: Angular beam divergence/Trap
velocity distributions}

Transverse velocity distributions put an additional constraint on the thin
SW lens performance and have been a limiting factor in Fresnel atom optics
and interferometry. For an atom beam, the angular divergence is equivalent
to the transverse velocity distribution of atoms in a cold trap, again
allowing a time domain treatment. The transverse and longitudinal beam
velocities are generally related linearly by the angle in the paraxial
limit. This is the case both for atomic beams collimated by apertures and
for beams that are laser-cooled in the transverse direction. Given that we
have already discussed chromatic aberration in detail, here we present
comparative focal parameters and densities which include either transverse
effects alone or simultaneous longitudinal and transverse velocity effects.
Only the narrow longitudinal velocity atom beam will be considered in detail
since this type of beam leads to a better focus.

We will look at two cases. For both cases the Fourier method suggests a
straightforward interpretation of the effects of a transverse velocity
distribution in terms of an inhomogeneous decay of the density spatial
harmonics. First, we treat the case where the beam is monochromatic as in
Sec. II and the transverse velocity distribution is either a uniform or
one-dimensional Maxwellian (thermal) distribution of transverse velocities.
The uniform distribution might apply to an atomic beam collimated by a slit
or a pair of slits while the thermal distribution might apply to a
laser-cooled beam or to atoms released from a laser-cooled trap. The
resulting focal parameters and densities for the slit-collimated and
laser-cooled atoms are similar qualitatively. Then, we consider a more
general beam case where the transverse velocity and longitudinal flux
distributions are {\it decoupled,} but the longitudinal velocity dependence
of the time {\it couples} the transverse integration to the longitudinal.
This applies to atomic beams that are laser cooled before interacting with
the SW lens. The theoretical treatment of each of these cases is facilitated
by the Fourier method.

Our first case is the monovelocity longitudinal beam ($v\rightarrow 0$) with
a transverse velocity distribution $P(U_{x})$. To unify the longitudinal and
transverse results, we start with Eq. (\ref{therm}) and take $W(U)=\delta
(U-U_{0}).$ This is equivalent to starting from Eq. (\ref{5a}) or (\ref{5b})
and taking $U=U_{0}$, $t=t_{0},$ and $\theta =\theta _{0}$. Assuming we are
describing the atomic beam rather than the trap, the atoms are again
detected, deposited, or used for lithography at $L=U_{0}t_{0}$. The
inclusion of an initial transverse velocity $U_{x}$ during the atom-field
interaction will lead to a Doppler shift of the field frequency. (Note that
the dimensionless transverse velocity scale that is consistent with the
dimensionless variables (\ref{a8}) is the single-photon recoil velocity, $%
V_{k}=\hbar q/2M$.) We rigorously account for the Doppler shift by taking $%
x\rightarrow x+U_{x}t$ and averaging the density over $P(U_{x})$, 
\begin{equation}
\rho \left( x,t_{0}\right) =\sum_{n=-\infty }^{\infty }e^{inx}J_{n}\left[
\theta _{0}\sin \left( nt_{0}\right) \right] \int
dU_{x}P(U_{x})e^{inU_{x}t_{0}}\text{.}  \label{doppler}
\end{equation}
The assumption is that the initial transverse density matrix is diagonal in
momentum space and can be described by a normalized velocity distribution $%
P(U_{x})$. In terms of the longitudinal velocity $U_{0}$ of the beam, the
transverse velocity can be written as $U_{x}=U_{0}\varphi ,$ where $\varphi $
is the propagation angle with respect to the $z$-axis.

The two transverse distributions of interest,  the uniform distribution $%
P_{1}(U_{x})$ and Maxwellian distribution $P_{2}(U_{x})$, take the following
form for $V_{x}\ll U_{0}$: 
\begin{mathletters}
\label{trans}
\begin{eqnarray}
1.\text{ }P_{1}(U_{x}) &=&(2V_{x})^{-1}\text{ for }U_{x}\in \lbrack
-V_{x},V_{x}]\text{,}  \label{trans1} \\
2.\text{ }P_{2}(U_{x}) &=&(\sqrt{\pi }V_{x})^{-1}\exp [-(U_{x}/V_{x})^{2}]%
\text{.}  \label{trans2}
\end{eqnarray}
Each can be written as a distribution over beam angles $\varphi $, if
desired, by defining the divergence angle $\varphi _{d}\equiv V_{x}/U_{0}$.
The first distribution might be formed when a thermal transverse
distribution, peaked at $U_{x}=0$ $(\varphi =0)$ with velocity width much
greater than $V_{k},$ is collimated by a pair of slits which select a small
range of propagation angles up to $\varphi _{d}$. This beam has a $rms$
transverse velocity spread of $V_{1}=V_{x}/\sqrt{3}$. The second
distribution might be formed by transverse laser cooling, with or without
previous slit collimation, and has a $rms$ velocity width of $V_{2}=V_{x}/%
\sqrt{2}$. (Technically, the laser cooling process can create a non-diagonal
density matrix with transverse momentum state coherences that are not
accounted for here but can be incorporated into a more general Fourier
result. We assume the distribution is narrow and diagonal.)

Fortuitously, both distributions again give closed-form Fourier coefficients
for the density. Inserting Eqs. (\ref{trans}) into Eq. (\ref{doppler}), one
finds 
\end{mathletters}
\begin{mathletters}
\label{transdens}
\begin{eqnarray}
1.\text{ }\rho \left( x,t_{0}\right)  &=&\sum_{n=-\infty }^{\infty }e^{inx}%
\frac{\sin [nV_{x}t_{0}]}{nV_{x}t_{0}}  \nonumber \\
&&\times J_{n}\left[ \theta _{0}\sin \left( nt_{0}\right) \right] ;
\label{transdens1} \\
2.\text{ }\rho \left( x,t_{0}\right)  &=&\sum_{n=-\infty }^{\infty }\exp
[inx-(nV_{x}t_{0}/2)^{2}]  \nonumber \\
&&\times J_{n}\left[ \theta _{0}\sin \left( nt_{0}\right) \right] \text{.}
\label{transdens2}
\end{eqnarray}
The interpretation is the same as for an atomic free induction decay
experiment\cite{12}. The harmonics of the total density undergo an
inhomogeneous decay as each velocity class $U_{x}$ evolves with its own
Doppler-shifted frequency. The $P_{1}$ result was recently explained in
detail as it is isomorphic to the evolution of a model, one-dimensional BEC
or a degenerate Fermi gas after interacting with a SW pulse\cite{Jayson1}. A
similar expression to the $P_{2}$ result was used recently to explain the
decay of periodic echoes\cite{12} and the decay of a quasiperiodic atomic
focusing and Talbot scheme\cite{411}. For both distributions the harmonic
decay functions of Eqs. (\ref{transdens}) depend on $nV_{x}\theta _{0}^{-1}$
near the focus, $t_{0}\sim \theta _{0}^{-1}.$ For $V_{x}\lesssim 1$ only
harmonics greater than $n_{\max }\simeq \theta _{0}$ undergo significant
damping, so it appears that the transverse velocity width can be up to the
order of the recoil velocity to avoid a significant decay of the high-order
Fourier components necessary for a highly-peaked focus. However, the
restriction $V_{x}\lesssim 1$ may be sufficient but not necessary. We have
found in the previous sections that the lens parameters scale with $\theta
_{0},$ and, therefore, in Sec. V we show that the restriction on $V_{x}$ is
somewhat relaxed.

The focal distance $t_{f}$, defined as the first maximum of Eq. (\ref
{transdens1}) or (\ref{transdens2}), moves closer to the lens for $V_{x}>0$
when compared to the focal distance for $V_{x}=0.$ This makes sense since
the transverse distribution causes a decay in the modulated density in time.
However, this shift of the focal distance is secondary in importance to the
changes in the peak density at the focus and the spot size. Moreover, as $%
\theta _{0}$ increases, the focal distance including the transverse velocity
spread converges rapidly to the ideal value near $t_{0}\simeq \theta
_{0}^{-1}$. We have calculated the peak density and spot size for selected
values of $V_{x}$ and compared them to the ideal case. In Fig. \ref{fig11}
the peak focal density, $\rho \left( 0,t_{f}\right) ,$ and spot size $w$ are
plotted versus $\theta _{0}$ for $V_{x}=0$ (ideal), $V_{1}=V_{2}=1$ (recoil
limit)$,$ and $V_{1}=V_{2}=2$ (twice the recoil limit). For each of the five
curves, the exact, different focal time at each value of $\theta _{0}$ is
used to calculate the parameters. For the same $rms$ velocity spread, the $%
P_{2}$ distribution gives a narrower, more peaked density spot than the $%
P_{1}$ distribution. It is significant that the percentage error in the peak
density and spot size is decreasing as $\theta _{0}$ increases. This result
is expected since inhomogeneous decay is less pronounced at earlier times, $%
t_{0}\sim \theta _{0}^{-1}$.

To treat the full problem of focusing an atomic beam with velocity
distributions more generally, we must account for the angular relation
between the longitudinal and transverse velocities. The atomic velocity in a
beam, ${\bf U}=U\hat{z}+U_{x}\hat{x},$ can be written ${\bf U}\simeq \left| 
{\bf U}\right| \hat{z}+\left| {\bf U}\right| \varphi \hat{x}$ for small
angles $\varphi ,$ where $\left| {\bf U}\right| \simeq U$ to lowest order in 
$\varphi $ in the paraxial approximation. Since atomic beams are inherently
formed by a process which truncates either the divergence angle or the
transverse velocity distribution, it is not a simple matter to describe them
theoretically. However, if the beam is localized around some large
longitudinal speed and then laser cooled in the transverse direction before
the SW lens, the longitudinal and transverse distributions are effectively
independent. We can model the total (normalized) flux probability $W({\bf U}%
)d{\bf U}$ as a product of the two independent distributions, $%
W_{lM}(U)P_{2}(U_{x})dUdU_{x},$ using the local Maxwellian longitudinal flux
distribution $W_{lM}(U),$ Eq. (\ref{172a}), and thermal transverse velocity
distribution $P_{2}(U_{x}),$ Eq. (\ref{trans2}).
\begin{figure}[tb!]
\centering
\begin{minipage}{8.0cm}
\epsfxsize= 8 cm \epsfysize= 7.34 cm \epsfbox{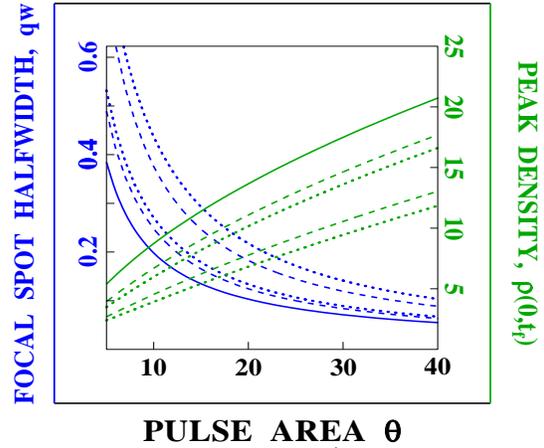}
\end{minipage}
\caption{Angular divergence (transverse velocity distribution) in atom
focusing, neglecting chromatic aberration. Exact pulse area dependence of
the lens parameters: Peak focal density $\protect\rho (0,t_{f})$ and spot
size $w$ (HWHM of the density profile at the focal plane). The solid curves
are for the ideal atom beam (with no angular divergence), the dotted curves
are for the uniform transverse distribution ($P_{1}$), and the dashed curves
are for the Maxwellian distribution ($P_{2}$). The middle pairs of curves,
closest to the ideal case, are for $V_{1}=V_{2}=1$, the $rms$ single-photon
recoil limit. The outer pairs are for $V_{1}=V_{2}=2$, twice the recoil
limit.}
\label{fig11}
\end{figure}

Since the time of flight, $t=L/U,$ to the detection plane, $L=U_{0}t_{0},$
depends on the longitudinal velocity, the Doppler phase $%
e^{inU_{x}t_{0}U_{0}/U}$ does as well. Inserting the integration of the
Doppler phase over the distribution $P_{2}(U_{x})$ into Eq. (\ref{therm}),
we find the velocity-averaged density to be 
\end{mathletters}
\begin{mathletters}
\label{densP}
\begin{eqnarray}
\bar{\rho}\left( x,t_{0}\right)  &=&\sum_{n=-\infty }^{\infty
}e^{inx}\int_{0}^{\infty }dUW\left( U\right) \exp
[-(nV_{x}t_{0}U_{0}/2U)^{2}]  \nonumber \\
&&\times \left\{ J_{n}\left[ \left( \theta _{0}U_{0}/U\right) \sin \left(
nt_{0}U_{0}/U\right) \right] \right\} .  \label{densP2b}
\end{eqnarray}
The Fourier component of the density now has an additional Gaussian
multiplier to integrate over longitudinal velocities. If the longitudinal
width $v$ is small but non-negligible and we want $V_{x}\lesssim 1$ (laser
cooling to the recoil limit), the argument of the Gaussian is on the order
of the argument of the sine function. As a result, we cannot neglect the
Gaussian decay in the integral by assuming $W\left( U\right) $ is sharply
peaked around $U_{0}$ and thus evaluating it at $U=U_{0}$ ($v\rightarrow 0$
limit). In Sec. V we treat two examples of experimental interest with this
expression, an atomic beam which is laser-cooled to the transverse recoil
limit, both with and without a longitudinal flux width $v.$ This is shown in
Fig. \ref{fig08} along with the ideal and purely longitudinally-broadened
cases.

\section{Discussion}

In this article we have developed a theory of atom focusing by standing wave
light fields in the thin lens regime. We have shown that the exact
analytical expressions for the Fourier components of the atomic density,
recently used in the theory of atom interferometry, are useful for numerical
calculations of the focusing effect and all its relevant lens parameters,
including the effects of chromatic aberration and angular divergence. We
call the Fourier technique Method 1.

Thin lens focusing becomes especially effective for a large field area $%
\theta $. When $\sqrt{\theta }\gg 1$ for the monovelocity case, using the
Fresnel-Kirchhoff diffraction integral, we can consider only the first
anharmonic term of the phase shift created by the SW light field when
expanding around the potential wells at $x^{\prime }=2\pi m$. This
procedure, labeled Method 2, leads to a manageable expression (\ref{a12})
for the atomic density distribution which can be used to evaluate the focal
parameters by numerical integration.

Within a small vicinity of the focal time, Eq. (\ref{a12}) simplifies
further to the asymptotic density profile (\ref{7}). This is Method 3. The
analytical correction to the focal time and analytical expressions for the
peak focal density, spot size, and depth of focus as functions of the pulse
area $\theta $ follow from this result. The last three lens parameters scale
as $\theta ^{1/2}$, $\theta ^{-3/4},$ and $\theta ^{-3/2}$ respectively,
while the focal time scales as $\theta ^{-1}\left( 1+1.27\theta
^{-\,^{1}/_{2}}\right) $ [see Eqs. (\ref{a14}), (\ref{a15}), (\ref{a16}),
and (\ref{a151})]. From Fig. \ref{fig03} one sees that in the range $%
0<\theta <40,$ the accuracy of the Method 3 asymptotic expressions for the
peak focal density, focal time, and spot size can be as small as 17\%, 1.9\%
and 18\% respectively. The asymptotic parameters obtained by Method 2 using
Eqs. (\ref{a12}) are generally more precise, giving accuracies for the same
parameters as small as 5.8\%, 0.24\% and 4\%, respectively.

The accuracy of the asymptotic result for the focal time $t_{f}$ is an order
of magnitude better than for other focusing parameters. Evidently, this is a
consequence of the fact that a finite value for the focal time, $t\sim
\theta ^{-1}$, comes out of the harmonic approximation for the lens
potential. The first anharmonic term then leads to a relative correction of
the order of $\theta ^{-1/2}$ [see Eq. (\ref{a14})], and higher-order
anharmonic corrections (not included in our consideration) should have
relative weights $\theta ^{-1}$ and larger inverse powers of $\theta $. As
for the other parameters, which can be derived only by including the first
anharmonic term, the next order corrections should have a larger relative
weight of $\theta ^{-1/2}$.

For example, the constant offsets of the asymptotic peak densities from the
exact result in Fig. \ref{fig03} can be explained as follows. A correction
to the asymptotic wave function at $x=0$ [Eq. (\ref{a13})] of relative
weight $\theta ^{-1/2}$ would result from including the next order spherical
aberration, the $x^{\prime 6}$ term in the expansion of $\cos (x^{\prime })$
in Eq. (\ref{a11a}). When the wave function is squared to form the density,
this relative correction of weight $\theta ^{-1/2}$ multiplies the
lowest-order peak density that grows as $\theta ^{1/2}$ to give a constant
term, which is automatically accounted for by the exact Fourier method.
Hence, the constant offset of the asymptotic peak density is not included in
our asymptotic expressions.

In this article we have also analyzed a new type of atom focusing which
arises when the SW field is resonant with the atomic transition. The density
profiles and focal parameters for the resonant lens are shown in Figs. \ref
{fig04} and \ref{fig05}, respectively. To gain insight into the different
interactions that can lead to focusing, we can consider the far-detuned and
resonant cases from a more general point of view than that taken in Secs. II
and III, still assuming that the field envelope is short enough to neglect
spontaneous emission during the pulse. When an atom interacts with a pulsed
light field, the states of the system can be decomposed into a set of
semiclassical dressed states of the atom plus field. If the atom-field
interaction is adiabatic or the pulse turns on instantaneously, these
dressed states are instantaneous eigenstates of the total Hamiltonian and
therefore undergo a phase modulation. When the light field modes have a
modulated intensity, the dressed state energies and, therefore, phase
evolutions are also spatially modulated. Focusing can occur near the
intensity extrema, which correspond to spatial minima of the dressed state
energies. If initially the bare atom was in a pure state such as the ground
state $|g\rangle $, the atomic wave function after the interaction is
generally a coherent admixture of the two dressed states, which have an
energy separation that is spatially modulated. Therefore, an interference
term in squaring the bare state amplitudes can lead to a Rabi-like
oscillation of the total density and its properties. This Rabi-like
oscillation can be seen in Eq. (\ref{a20a}) and Fig. \ref{fig05} for a
resonant SW field. Nonadiabatic, detuned atom-field couplings can lead to a
similar effect. The evolution of the pure and dressed states in this regime
has been considered previously without center-of-mass spatial effects\cite
{Lu}. For the far detuned standing wave $\left( \left| \Delta \right| \gg
\left| \chi \right| \right) $ or an adiabatic turn-on of the field, the
ground state evolves into only one of the dressed states while the other
dressed state has a negligible amplitude. As a result, one does not observe
any interference or field dependent oscillation in this case.

Up to this point in the discussion, we have reviewed the ideal situation of
the ideal atomic beam having no angular divergence and no longitudinal
velocity distribution. Our calculations in the thin lens regime have shown
that focusing results in a relatively large peak focal density. For the most
part, this result is in contrast to previous experiments carried out in the
thick lens regime, where it proved difficult to achieve such large density
peaks. The exceptions are certain experiments in Na at $\lambda =589\,nm$ on
the $3S_{1/2},\,F=2\rightarrow 3P_{3/2},\,F^{\prime }=3$ transition\cite
{6p,61}. In these thick lens experiments for a thermal beam having thermal
velocity $V\simeq \,8.6\times 10^{4}\,cm/s,$ the spot size resolution and
contrast were $w=10\,nm,$ $c(t_{f})=10$\cite{6p} and $w=6.5\,nm,$ $%
c(t_{f})=6 $\cite{61}, respectively. As an exercise, we can compare that
experimental data using a thermal beam and thick SW lens with our results
using a ideal atomic beam and thin lens. From Fig. \ref{fig03} or an exact
calculation, a pulse area of $\theta \simeq 14.86$ ($9.37)$ is needed for a
resolution of $qw=0.139$ ($qw=0.139$), or $w=6.5$ $nm$ ($10$ $nm$) in
sodium. If we set the beam velocity $U$ equal to $8.6\times 10^{4}\,cm/s,$
then focusing for $\theta \simeq 14.86$ occurs at a distance $%
L_{f}=Ut_{f}=127\,\mu m.$ The atomic density profile at this distance is
shown with the thin solid line in Fig. \ref{fig08}. The contrast at the
focus, defined by Eq. (\ref{contrast}) as the ratio of focal atomic
densities at $qx=0$ and $qx=\pi $ (not shown in Fig. \ref{fig08}), is $%
c(t_{f})=27.2.$

We can determine the standing wave power needed to achieve a pulse area of $%
\theta =14.86$ for the two-level system. This system can be realized in Na
using optical pumping to the ground state sublevel having magnetic quantum
number $m_{F}=\pm 2.$ A circularly polarized standing wave drives only the $%
3S_{1/2},\,\,F=2,\,m_{F}=\pm 2\rightarrow 3P_{3/2},\,\,F^{\prime
}=3,\,m_{F^{\prime }}=\pm 3$ transition which has the dipole moment matrix
element $\left| \mu \right| =6.37\times 10^{-18}$ $esu\cdot cm$. To achieve
the intensity necessary for atomic focusing, the SW field must be formed by
focused laser beams. In the experiments of Refs.\cite{6p,61}, the fields
have been focused to a circular spot having radius $\sigma _{z}\simeq
29\,\mu m$\cite{thin}. For experiments in the thin lens regime, it is
sufficient to focus the field only in the $z$-direction, along which the
atomic beam propagates. We assume that the laser field intensity has a
Gaussian profile in the $\left( y,z\right) $ plane, 
\end{mathletters}
\begin{equation}
\left| E\left( y,z\right) \right| ^{2}=\left| E\right| ^{2}\exp \left(
-2y^{2}/\sigma _{y}^{2}-2z^{2}/\sigma _{z}^{2}\right) ,  \label{25}
\end{equation}
where $\sigma _{z}$ and $\sigma _{y}$ are field radii along the $z$- and $y$%
-axes, respectively (i.e., the directions perpendicular to the laser beam
propagation along the $x$ axis). For homogeneous atomic focusing into a set
of lines, one should choose $\sigma _{y}$ to be larger than the atomic beam
radius. From the data in Refs.\cite{6p,61}, $\sigma _{y}$ should be as large
as $0.5\,cm.$ The general field mode (\ref{25}) can be created using
cylindrical optical lenses. The far-detuned pulse area $\theta $ (\ref{2})
can be re-expressed through the constituent traveling wave field powers of
Eqs. (\ref{1}) and (\ref{25}), $P=c\left| E\right| ^{2}\sigma _{z}\sigma
_{y}/16$, as $\theta =-\sqrt{\pi }2^{7/2}\left| \mu \right| ^{2}P/(\hbar
^{2}c\sigma _{y}\Delta U).$ For further estimates we assume that $\sigma
_{z}=29\,\mu m$, $\sigma _{y}=0.5\,cm,$ and $\Delta =-2\pi \cdot
(1.71\,GHz). $ Then, for a field area of $\theta =14.86,$ the required laser
power is $P=1.89\,mW.$ Even this power level is four times less than the
power used in the experiment of Ref.\cite{61}. To compare directly to the
circular focus of the field in Ref.\cite{61}, we must take $\sigma
_{y}=29\,\mu m,$ giving us a required power of $P=11.0\,\mu W.$ These
relatively low power levels are sufficient for our consideration of thin
lens focusing.

We can now discuss the influence of chromatic aberration assuming no angular
divergence. A longitudinal velocity distribution in the atomic beam leads to
a degradation of the focus. But even for the thermal beam, the normalized
distribution width $v$ is still less than $1$ [see Eq. (\ref{221})], and we
have shown that the order of magnitude of the focusing parameters is the
same as for the ideal beam ($v=0$). For narrow flux distributions, $v\ll 1,$
the lowest-order corrections to the atom density behave as $v^{2}$ [see Eqs.
(\ref{20})] if $\theta \gtrsim 1$. This is a simple consequence of the fact
that terms linear in the deviation from the average atomic flux vanish after
averaging over velocities for any symmetric, or asymmetric, distribution. In
particular, this principle applies to the analytical Fourier components of
the density multiplied by the narrow, local Maxwellian flux distribution $%
W_{lM}(U)$, Eq. (\ref{therm}), for small $v$.

In Fig. \ref{fig07} we show the exact (using $W_{lM}(U)$) and approximate
reductions of the density at the ideal beam focal time, $\bar{\rho}%
(0,t_{0,f})/\rho \left( 0,t_{0,f}\right) $\cite{peaks}$,$ as functions of $v$
for two values of $\theta _{0}$, $14.86$ and $9.37$. Again, these choices
for $\theta _{0}$ are the pulse areas required to achieve $w_{0}=6.5$ and $%
10 $ $nm$ spot sizes \cite{6p,61}, respectively, using ideal sodium beams.
The quadratic dependence, consistent with Eqs. (\ref{21c}), is evident for
small $v$, but the exact density peak actually reduces more slowly
(linearly) at larger values of $v$ where the small $v$ approximation breaks
down. At this point the expansions of the Bessel functions of Eq. (\ref
{therm}) near $t_{0,f}\approx \theta _{0}^{-1}$ in terms of $u$ are invalid
since $n_{\max }\left\langle \left| u\right| \right\rangle \gtrsim 1$, or
equivalently $v\sim \theta _{0}^{-1}$: a term proportional to $\left|
u\right| $ would appear in an expansion around the average velocity. We take
this as evidence that near this value of $v$ for a fixed $\theta _{0},$ the
focal properties go from being dominated by spherical aberration to a regime
where a more complicated combination of spherical and chromatic aberrations
is important.

To test this idea, we have also examined the dependence of the spot size
increase, $w/w_{0},$ on $v$ for $\theta _{0}=14.86$ and $9.37$ using the
exact density for $W_{lM}(U)$. In Fig. \ref{fig07} the spot size's quadratic
dependence for small $v$ turns into a linear dependence near the same value
of $v$ that the peak density reduction deviates from the approximate theory.
We see that the chromatic aberration has a smaller effect on the spot size
than on the peak density. Note that from a ray tracing argument for
far-field focusing through an effective parabolic lens aperture of full
width $\sim \lambda /2$, a strong, $\theta $-independent, linear dependence
of $w\propto \pi v$ (or equivalently $\left\langle \left| u\right|
\right\rangle $) would be expected for a focus dominated by chromatic
aberration in the absence of spherical aberration\cite{84}. However, this is
not the case for the SW field lens, where spherical aberration determines
the limiting resolution up to large values of $\theta _{0}$. Thus, the
chromatic aberration we find is less severe and depends strongly on $\theta
_{0}$. We have not explored this $\theta _{0}$ dependence further. However,
from Fig. \ref{fig07} we deduce that if we fix $v$ (and by implication the
atomic beam properties), a threshold pulse area, $\theta _{th}\sim v^{-1}$,
must exist for each $v$ to mark the breakdown of the small $v$ expansion.
For $\theta _{0}\gtrsim \theta _{th}$ the effects of chromatic aberration
should be calculated exactly.

To further demonstrate the small $v$ results for a narrow longitudinal
distribution, we can define $\alpha $ to be the percentage reduction of the
peak focal density as a function of $\theta _{0}$ and $v$. The flux
distribution width for a given $\theta _{0}$ and $\alpha $ is therefore $v=%
\sqrt{\alpha \rho \left( 0,t_{0,f}\right) /\rho _{1}\left( 0,t_{0,f}\right) }
$ from Eq. (\ref{21c}). From Fig. \ref{fig06} or \ref{fig07} for $\theta
_{0}=14.86$, to limit the peak density reduction to $\alpha =10\%$, we can
estimate that a normalized flux width of $v\simeq 0.086$ would be needed.
(For the local Maxwellian distribution, $W_{lM}(U),$ a flux width of $%
v\simeq 0.091$ is found to be sufficient by an exact calculation with Eq. (%
\ref{therm}).) Using $v\simeq 0.086$ to calculate the other parameters
approximately from Eqs. (\ref{21b}) and (\ref{21e}), the percentage
corrections to the focal distance and spot size are even less than $\alpha ,$
$3.9\%$ and $1.3\%,$ respectively. A longitudinal width, $v\simeq 0.086,$ is
typical of the supersonic beams produced by seeding an inert (noble) gas
supersonic expansion with sodium\cite{900}. While the mean longitudinal
velocity of a beam produced recently from a BEC of sodium atoms by Bragg
scattering\cite{13} was only $2\hbar k/M,$ or $6$ $cm/s,$ the $rms$
longitudinal and transverse velocity widths were only approximately $%
0.16\hbar k/M$ and $0.30\hbar k/M$, respectively, giving $v\simeq 0.11$ and
justifying the paraxial approximation. In these experiments, mean
longitudinal velocities up to $11.9$ $\hbar k/M,$ or $35$ $cm/s$, with
similar widths were also achieved using higher-order Bragg scattering,
implying $v\ll 0.1$. Since experiments producing atomic beams of this type
are in their infancy, we expect that achieving relative velocity widths of $%
v\ll 0.1$ will shortly become routine.

In Fig. \ref{fig08} the atomic distribution at the focus $\bar{\rho}\left(
x,t_{f}\right) $ (dotted line), as calculated exactly using $W_{lM}(U)$ in
Eq. (\ref{therm}) for $\theta _{0}=14.86$ and our extreme case from Fig. \ref
{fig07} of $v\simeq 0.194$ ($V/\widetilde{U}_{0}=0.2$)$,$ is compared to the
ideal case (thin solid line). While the reduction of the peak density from $%
11.4$ to $7.81$ is a significant $31\%$ on a scale where $\sim 0.5$ is the
background at the focus, the half-width defining the spot size is broadened
by only $5\%$ (or $0.33$ $nm$ in sodium), and the coherent oscillations of
the density along $x$ are still prevalent. This suggests that even for $%
v\approx 0.2$ and a focal peak reduced by chromatic aberration, a high
resolution, large contrast focus is possible with the thin SW lens. In
addition, using the flux distribution $W_{lM}(u)$ with $v=0.194,$ we have
also calculated that the field area, $\theta _{0}=15.7,$ is required to
restore a spot size of $qw=0.139$, or $6.5$ $nm$ in sodium. This corresponds
to the laser power, $P=2.0\,mW,$ for $\sigma _{z}=29\,\mu m$, $\sigma
_{y}=0.5$ $cm,$ and $\Delta =-2\pi \cdot (1.71\,GHz)$.

We have also considered the focusing of a thermal atomic beam with flux $%
W_{M}(U)$. The results of the calculations are graphed in Figs. \ref{fig09}
and \ref{fig10}. For this part of the paper only, we changed the definition
of the focal distance to be one where the contrast (\ref{contrast}), and not
the density, is optimized. The contrast, $c\left( t_{0}\right) =\bar{\rho}%
\left( 0,t_{0}\right) /\bar{\rho}\left( \pi ,t_{0}\right) ,$ for $\theta
_{0}=23.0$ at a distance $L$ from the SW field $\left( t_{0}=L/V\right) $ is
shown in Fig. \ref{fig09}. We see that $c\left( t_{0}\right) $ contains
three local maxima of approximately equal weight near the focus. They arise
as a result of\ the time-oscillations of the flux-averaged, background atom
density near $t_{0}=t_{f}$, $\bar{\rho}\left( \pi ,t_{0}\right) $. These
contrast maxima ''compete'' with one another in some sense as different
values of the pulse area correspond to different velocity classes and
therefore different focal distances. This effect results in the
discontinuities in the pulse area dependences of the lens parameters seen in
Fig. \ref{fig10}.

For the longitudinal thermal beam and $\theta _{0}=23.0$, the focal contrast
is equal to $c\left( t_{f}\right) =8.03$ while the spot size or spatial
resolution is again $qw=0.139.$ For the same experimental parameters as
above, $\sigma _{z}=29\,\mu m$, $\sigma _{y}=0.5$ $cm,$ and $\Delta =-2\pi
\cdot (1.71\,GHz)$, a laser field power of $P=2.9\,mW$ is needed to produce
a pulse area of $\theta _{0}=23.0$.

Finally, the atomic beam angular divergence has a crucial impact on the
focus, independent of the chromatic aberration. For a beam with no chromatic
aberration, $U=U_{0}$, if atoms move at an angle $\varphi $ with respect to
the $z$-axis, the density profile displaces a distance $\delta x\sim
U_{0}\varphi t_{f}\simeq U_{x}t_{f}$ along the $x$ axis. This displacement
should ideally be much smaller than the fundamental spot size $w$ that
accounts for spherical aberration. We can use the previously generated,
exact spot sizes and focal times for each $\theta _{0}$ to estimate the
maximum allowed $rms$ transverse velocity from this argument, $%
V_{rms}^{(f)}\approx w/t_{f}$. This corresponds to an angular divergence $%
\varphi _{f}=V_{rms}^{(f)}/U_{0}\approx w/(t_{f}U_{0})$. We expect that
allowing $V_{rms}$ to be $V_{rms}^{(f)}$ will at most double the spot size.

Alternatively, from Eqs. (\ref{a14}) and (\ref{a16}) for dimensionless $%
t_{f} $ and $w$, the asymptotic restriction on the angular divergence, $%
\varphi \lesssim \varphi _{f},$ is 
\begin{equation}
\varphi _{f}\sim 0.744\frac{\hbar q}{2MU_{0}}\theta
_{0}^{1/4}(1-1.27\theta _{0}^{-1/2})\sim 0.744\frac{V_{k}}{U_{0}}\theta
_{0}^{1/4}.  \label{26}
\end{equation}
It is interesting to note that the asymptotic focusing restriction is $\sim
\theta ^{1/4}$ times less severe than the recoil limit, 
\begin{equation}
\varphi \ll \varphi _{T}\sim \frac{\hbar q}{2MU_{0}}\equiv \frac{%
\lambda _{dB}}{\lambda },  \label{27}
\end{equation}
which is the condition required to observe atom interference in the single
interaction region geometry (for example, the atomic Talbot effect\cite
{86a,rev,23,24}). Condition (\ref{27}) arises from the requirement that
atoms moving at an angle $\varphi $ not be displaced more than $\lambda /2$
at the Talbot distance $L_{T}\sim \lambda ^{2}/2\lambda _{dB}$. Even
though effective focusing requires the displacement to be $\theta ^{3/4}$
times smaller asymptotically (since $qw\propto \theta ^{-3/4}$), this
displacement occurs at a distance that is $\theta $-times shorter than the
Talbot distance (since $L_{f}=U_{0}t_{f}\sim L_{T}/\theta _{0}$).

This argument is consistent with Eqs. (\ref{transdens}) and Fig. \ref{fig11}%
. This figure suggests that small spot sizes with sharp density peaks can be
achieved for laser powers in the microwatts to milliwatts range even for
transverse distribution widths larger than the recoil limit. (In Fig. \ref
{fig11}, note that $qw=0.3$ corresponds to a $14$ $nm$ spot size in sodium.)
For example, taking $\theta _{0}=40$ and remembering that the
slit-collimated beam has a $rms$ transverse velocity $V_{1}$ ($V_{x}\sqrt{3}$%
) in units of $V_{k}$, the exact spot size for $V_{1}=0$ is $qw=0.0577$ from
Eq. (\ref{transdens1}). Now, assuming the divergence is the maximum allowed,
we set $V_{1}=w/t_{f}=1.883$ for $\theta _{0}=40$ and find that the spot
size is broadened to $0.100$ ($4.69$ $nm$ in sodium), or $73\%$. The spot
size has less than doubled for a transverse width almost twice the recoil
limit. Similarly, the asymptotic predictions for $\theta _{0}=40$ give an
increase from a spot size of $qw=0.0468$ (not $0.0577$) for $V_{1}=0$ to a
spot size of $0.0826$ for $V_{1}=U_{0}\varphi _{f}=1.495$ from Eq. (\ref{26}%
) in Eq. (\ref{transdens1}), a $76\%$ increase. The broadening is even less
severe for Maxwellian transverse distributions.

We now take a case of experimental interest to demonstrate the combined
effects of transverse and longitudinal broadening on the focal density. The
pulse area, $\theta _{0}=14.86$, is used again. We show results for the $%
P_{2}$ (Maxwellian) transverse distribution, Eq. (\ref{trans2}), since laser
cooling has been shown to improve the focus in thick lens experiments\cite
{6,6p,61}. In Fig. (\ref{fig08}) we plot the density at the focus $\rho
\left( x,t_{f}\right) $ (dashed line) from Eq. (\ref{transdens2}) for a
monochromatic beam ($v\rightarrow 0$) cooled to the $rms$ single-photon
recoil limit, $V_{2}=1$ $(V_{x}=\sqrt{2}).$ In addition, we show the atomic
focal density (lower solid line) from Eq. (\ref{densP2b}) for $\theta
_{0}=14.86$, $v=0.194$ ($V/\widetilde{U}_{0}=0.2$ in $W_{lM}(U)$), and $%
V_{2}=1$. The latter curve combines the attributes of the other curves in
this figure to provide a real picture of the focus. The following table
summarizes the results for Fig. \ref{fig08}: 

Table 2. Focusing of an atomic beam for $\theta _{0}=14.86.$ A
velocity-averaged spot of $qw=0.190$ corresponds to $8.91$ $nm$ in sodium,
for which $V_{k}\simeq 2.95$ $cm/s$ and $\omega _{q}\simeq 6.29\times 10^{6}$
$rad/s$.
\[
\begin{tabular}{|c|c|c|c|c|c|c|c|}
\hline
$
\begin{array}{c}
\text{Atomic} \\ 
\text{Beam}
\end{array}
$ & $v$ & $\frac{V_{x}}{V_{k}}$ & $\omega _{q}t_{f}$ & $\rho \left(
0,t_{f}\right) $ & $qw$ & $\rho \left( \pi ,t_{f}\right) $ & $c(t_{f})$ \\ 
\hline
Ideal & $0$ & $0$ & $.0930$ & $11.4$ & $.139$ & $.420$ & $27.2$ \\ \hline
$
\begin{array}{c}
\text{Chromatic} \\ 
\text{Aberration}
\end{array}
$ & $.194$ & $0$ & $.0856$ & $8.06$ & $.146$ & $.438$ & $18.4$ \\ \hline
$
\begin{array}{c}
\text{Angular} \\ 
\text{Divergence}
\end{array}
$  & $0$ & $\sqrt{2}$ & $.0893$ & $9.06$ & $.169$ & $.430$ & $21.1$ \\ \hline
Combined & $.194$ & $\sqrt{2}$ & $.0814$ & $6.64$ & $.190$ & $.450$ & $14.8$
\\ \hline
\end{tabular}
\]

Our work suggests emphatically that the thin SW lens can focus atoms
effectively if the angular divergence is cooled near the recoil limit, even
in the presence of strong chromatic aberration. This makes physical sense if
one considers that the longitudinal velocity average over slightly different
focal regions for each velocity subclass is a slowly-varying integral over
the Fourier {\it amplitudes} when compared to the more sensitive transverse
velocity integral, an average over the Doppler {\it phases}. Atomic beams of
the type considered throughout this article can be made in the laboratory
with current technology.

In addition to the experimental possibilities for standard atomic beams that
limit angular divergence by slit collimation or transverse laser cooling,
recent experiments that use higher-order Bragg diffraction of a BEC to form
atomic beams with transverse and longitudinal $rms$ velocities smaller than
the recoil velocity\cite{13} offer a promising avenue to observe and
characterize thin lens focusing effects. We would even argue that focal
patterns nearly identical to those described in this paper have already been
achieved but not recognized in recent condensate experiments on the
so-called momentum space Talbot effect\cite{newnist}. In these experiments a
cloud of cold atoms from a condensate interacts with two off-resonant SW
pulses, separated by a variable time delay. The detection scheme is
insensitive to Fresnel effects as the researchers image the far-field
diffraction pattern and therefore the atomic momentum distribution. However,
the accurate fit of the data in that experiment to our Raman-Nath theory
suggests that the thin lens focusing effect is occurring after the first
pulse along the SW grating direction. A detection scheme sensitive to the
Fresnel density pattern may be necessary to image the focusing atoms in the
cloud.

\acknowledgments

J.L.C. is indebted to Prof. Tycho Sleator and the NYU Physics Department for
providing him with the Visiting Scholar appointment during which this work
was completed. This work is supported by the National Science Foundation
under Grants No. PHY-9414020 and PHY-9800981, by the U.S. Army Research
Office under Grant No. DAAG55-97-0113 and AASERT No. DAAH04-96-0160, and by
the University of Michigan Rackham predoctoral fellowship.

\end{multicols}

\begin{references}
\bibitem{peaks}  Note that by fixing $t_{0,f}$ we are not actually using the
time $\bar{t}_{0,f}$ where the function $\bar{\rho}(0,t_{0})$ is a maximum
when evaluating $\bar{\rho}(0,t_{0,f})/\rho \left( 0,t_{0,f}\right) $. The
correction is small even for the largest value of $v$ plotted.
\bibitem{thin}  For this field area the Raman-Nath parameter (\ref{RN}) is
relatively small, $\theta \omega _{q}\tau /3=0.21$.
\end{references}
\end{document}